\documentclass[sigconf,screen]{acmart}

\copyrightyear{2025}
\acmYear{2025}
\setcopyright{cc}
\setcctype{by}
\acmConference[FSE Companion '25]{33rd ACM International Conference on the Foundations of Software Engineering}{June 23--28, 2025}{Trondheim, Norway}
\acmBooktitle{33rd ACM International Conference on the Foundations of Software Engineering (FSE Companion '25), June 23--28, 2025, Trondheim, Norway}\acmDOI{10.1145/3696630.3728541}
\acmISBN{979-8-4007-1276-0/2025/06}

\usepackage{algorithmic}
\usepackage{graphicx}
\usepackage{textcomp}

\PassOptionsToPackage{hyphens}{url}
\usepackage[nameinlink,capitalize]{cleveref}
\usepackage{enumitem}
\usepackage{etoolbox}
\usepackage{microtype}
\AtBeginEnvironment{verbatim}{\microtypesetup{activate=false}}
\usepackage{relsize}
\usepackage{booktabs}
\usepackage{multirow}
\usepackage{bm}
\usepackage{pifont}
\usepackage{float}

\crefname{appendix}{App.}{Apps.}

\usepackage{caption}
\usepackage{subcaption}
\usepackage{balance}

\usepackage{listings}

\definecolor{comment-green}{HTML}{609060}
\definecolor{starcomment-green}{HTML}{008000}
\definecolor{string-red}{HTML}{A31515}
\definecolor{keyword-blue}{HTML}{0000FF}
\definecolor{control-magenta}{HTML}{AF00DB}
\definecolor{function-brown}{HTML}{795E26}
\definecolor{type-teal}{HTML}{267F99}
\definecolor{number-teal}{HTML}{098658}
\definecolor{bracket-blue}{HTML}{0431FA}

\lstdefinestyle{CommonStyle}{
    basicstyle=\fontencoding{T1}\ttfamily\scriptsize,
    upquote=true,  
    frame=lines,  
    framesep=1pt,
    rulecolor=\color{gray!50},
    aboveskip=0pt,
    belowskip=-3pt,
    breaklines,
    breakatwhitespace=true,
    postbreak=\mbox{\textcolor{red}{\ensuremath{\hookrightarrow}\space}},
    showstringspaces=false,
    commentstyle=\color{comment-green},
    stringstyle=\color{string-red},
    keywordstyle=\color{keyword-blue},
}

\lstdefinestyle{MyPython}{
    language=python,
    style=CommonStyle,
    morestring=[s]{f'}{'},
    morestring=[s]{f"}{"},
    morecomment=[l][\color{starcomment-green}]{\#*},  
    keywords={and,class,def,False,global,is,lambda,None,nonlocal,not,or,self,True},
    keywords=[2]{as,assert,break,continue,del,elif,else,except,finally,for,from,if,import,in,pass,raise,return,try,while,with,yield},
    keywords=[3]{abs,aiter,all,anext,any,ascii,bin,breakpoint,callable,chr,compile,delattr,dir,divmod,enumerate,eval,exec,filter,format,getattr,globals,hasattr,hash,help,hex,id,input,isinstance,issubclass,iter,len,locals,map,max,memoryview,min,next,oct,open,ord,pow,print,range,repr,reversed,round,setattr,slice,sorted,sum,vars,zip,__import__},  
    keywords=[4]{classmethod,property,staticmethod,__init__,nearest_neighbor_tour},  
    keywords=[5]{bool,bytes,bytearray,complex,dict,float,frozenset,int,list,object,set,str,super,tuple,type,ValueError,Exception},  
    keywordstyle=[2]\color{control-magenta},
    keywordstyle=[3]\color{function-brown},
    keywordstyle=[4]\color{function-brown},
    keywordstyle=[5]\color{type-teal},
    literate=*{0}{{\textcolor{number-teal}{0}}}{1}%
        {1}{{\textcolor{number-teal}{1}}}{1}%
        {2}{{\textcolor{number-teal}{2}}}{1}%
        {3}{{\textcolor{number-teal}{3}}}{1}%
        {4}{{\textcolor{number-teal}{4}}}{1}%
        {5}{{\textcolor{number-teal}{5}}}{1}%
        {6}{{\textcolor{number-teal}{6}}}{1}%
        {7}{{\textcolor{number-teal}{7}}}{1}%
        {8}{{\textcolor{number-teal}{8}}}{1}%
        {9}{{\textcolor{number-teal}{9}}}{1}%
        {.0}{{\textcolor{number-teal}{.0}}}{2}
        {.1}{{\textcolor{number-teal}{.1}}}{2}
        {.2}{{\textcolor{number-teal}{.2}}}{2}%
        {.3}{{\textcolor{number-teal}{.3}}}{2}%
        {.4}{{\textcolor{number-teal}{.4}}}{2}%
        {.5}{{\textcolor{number-teal}{.5}}}{2}%
        {.6}{{\textcolor{number-teal}{.6}}}{2}%
        {.7}{{\textcolor{number-teal}{.7}}}{2}%
        {.8}{{\textcolor{number-teal}{.8}}}{2}%
        {.9}{{\textcolor{number-teal}{.9}}}{2}%
        {(}{{\textcolor{bracket-blue}{(}}}{1}%
        {)}{{\textcolor{bracket-blue}{)}}}{1}%
        {[}{{\textcolor{bracket-blue}{[}}}{1}%
        {]}{{\textcolor{bracket-blue}{]}}}{1}%
        {\{}{{\textcolor{bracket-blue}{\{}}}{1}%
        {\}}{{\textcolor{bracket-blue}{\}}}}{1}%
        ,%
}

\lstdefinestyle{MyJava}{
    language=java,
    style=CommonStyle,
    deletekeywords={if, else, for, while, do, switch, case, try, catch, finally, throw, return, new, break, continue},
    keywords=[2]{if, else, for, while, do, switch, case, try, catch, finally, throw, return, new, break, continue},
    keywords=[5]{Exception, ArrayList, String, Long, Integer, Boolean, Double, Float},  
    keywordstyle=[2]\color{control-magenta},
    keywordstyle=[5]\color{type-teal},
    literate=*{(}{{\textcolor{bracket-blue}{(}}}{1}%
        {)}{{\textcolor{bracket-blue}{)}}}{1}%
        {[}{{\textcolor{bracket-blue}{[}}}{1}%
        {]}{{\textcolor{bracket-blue}{]}}}{1}%
        {\{}{{\textcolor{bracket-blue}{\{}}}{1}%
        {\}}{{\textcolor{bracket-blue}{\}}}}{1}%
        ,%
}

\makeatletter
\patchcmd{\lsthk@SelectCharTable}{%
  \lst@ifbreaklines\lst@Def{`)}{\lst@breakProcessOther)}\fi
}{%
}{
}{
}
\makeatother

\makeatletter
\lst@AddToHook{OnEmptyLine}{\vspace{\dimexpr-\baselineskip+5pt}}
\makeatother

\usepackage{adjustbox}
\newcommand{\llmtext}[1]{``\texttt{\smaller\color{blue!50!black}{#1}}''}
\newcommand{\commentcode}[1]{\adjustbox{margin=0.5ex,bgcolor=gray!6}{\texttt{\smaller\color{green!30!black}{#1}}}}
\newcommand{\code}[1]{\protect\adjustbox{margin=0.5ex,bgcolor=gray!6}{\texttt{\smaller\color{blue!50!black}{#1}}}}
\newcommand{\circled}[1]{{\larger\textcircled{\smaller[2]{#1}}}}
\newcommand{\highlight}[1]{\adjustbox{margin=0.5ex,bgcolor=yellow!20}{#1}}

\newcommand{\xmark}{\text{\ding{55}}}


\begin{document}

\title{Natural Language Outlines for Code:\texorpdfstring{\\}{ }Literate Programming in the LLM Era}
\renewcommand{\shorttitle}{Natural Language Outlines for Code: Literate Programming in the LLM Era}

\author{%
Kensen Shi\text{*},
Deniz Altınbüken,
Saswat Anand,
Mihai Christodorescu,
Katja Grünwedel,
\mbox{Alexa Koenings},
Sai Naidu,
Anurag Pathak,
Marc Rasi,
Fredde Ribeiro,
Brandon Ruffin,
\mbox{Siddhant Sanyam},
Maxim Tabachnyk,
Sara Toth,
Roy Tu,
Tobias Welp,
Pengcheng Yin,
\mbox{Manzil Zaheer},
Satish Chandra,
Charles Sutton}
\affiliation{\institution{Google}\country{}}
\renewcommand{\shortauthors}{Shi et al.}

\thanks{\text{*}Corresponding author: Kensen Shi, kshi@google.com. Authors are listed alphabetically except for the first and last two authors. See \cref{app:authors} for author contributions.}

\begin{abstract}
We propose using \emph{natural language outlines} as a novel modality and interaction surface for providing AI assistance to developers throughout the software development process. An NL outline for a code function comprises multiple statements written in concise prose, which partition the code and summarize its main ideas in the style of literate programming. Crucially, we find that modern LLMs can generate accurate and high-quality NL outlines in practice. Moreover, NL outlines enable a bidirectional sync between code and NL, where a developer can change either code or NL and have the LLM automatically update the other. We discuss many use cases for NL outlines: they can accelerate understanding and navigation of code and diffs, simplify code maintenance, augment code search, steer code generation, and more. We then propose and compare multiple LLM prompting techniques for generating outlines and ask professional developers to judge outline quality. Finally, we present two case studies applying NL outlines toward code review and malware detection.
\end{abstract}

\maketitle

\section{Introduction}
\label{sec:intro}

Outlining a complex document is a standard way to improve its organization and accessibility. A \emph{table of contents} presents organizational notes separately from the document's content, but these notes may also be interleaved with the content as chapter or section headings. Modern examples of outlines include the ``Contents'' sidebar on Wikipedia pages listing the section headings, and for programming, the ``Outline view'' in VS Code and the ``Symbols pane'' on GitHub list the classes and functions in a code file.
These outlines also assist navigation by jumping to the corresponding part of the file when clicked.
Advances in large language models (LLMs) such as Gemini~\cite{GEMINI-1.0,GEMINI-1.5} and \mbox{GPT-4}~\cite{GPT-4} raise new opportunities to automatically generate outlines of software artifacts in reimagined ways.
After all, large software projects are incredibly complex, with one study finding that developers spend 70\% of their time on program comprehension~\cite{DEV-TIME}.

Summarization is another way to accelerate understanding of a document. Scientific papers have an \emph{abstract}, and business documents may have an \emph{executive summary}, to help readers quickly understand the key points without reading the entire text.
Even if one does read the entire text, reading the summary beforehand may still improve understanding~\cite{ABSTRACTS,PRIMING}. Recent works apply machine learning and LLMs toward automatic text summarization~\cite{ATS} and code summarization~\cite{CODE-SUMMARIZATION-survey-2022,CODE-SUMMARIZATION-survey-2019,hu2022practitioners,CODE-SUMMARIZATION-2022,CODE-SUMMARIZATION-2020,ahmed2024automatic,DOCUMENTATION-GENERATION-2024}, with many summary styles for different purposes and audiences~\cite{MULTI-INTENT-COMMENT,CODE-SUMMARIZATION-2024,mu2023developer,COMMENT-CATEGORIES}. For example, upon straightforwardly asking Gemini or ChatGPT to summarize a piece of code, one usually receives long paragraphs explaining each aspect of the code in detail, often totaling more text than the code itself. Such explanations are useful for learning but are not optimized for helping an experienced developer work more quickly. Instead, we desire a concise summary that efficiently conveys the key ideas, with clear alignment between the summary and the code.

Another common way to improve the clarity of code is to incorporate natural language (NL). Writing docstrings, comments, helpful error messages, and descriptive identifier names are well-known aspects of good coding style~\cite{PYTHON-STYLE}. In computational notebooks or other forms of \emph{literate programming}~\cite{LITERATE}, code and explanatory NL are interleaved in a file that is both executable and self-documenting. NL is generally easier for humans to read than code, so well-written NL can help developers understand the associated code more quickly.

\begin{figure*}
    \centering
    \includegraphics[width=\textwidth,trim={3mm 3mm 3mm 3mm},clip]{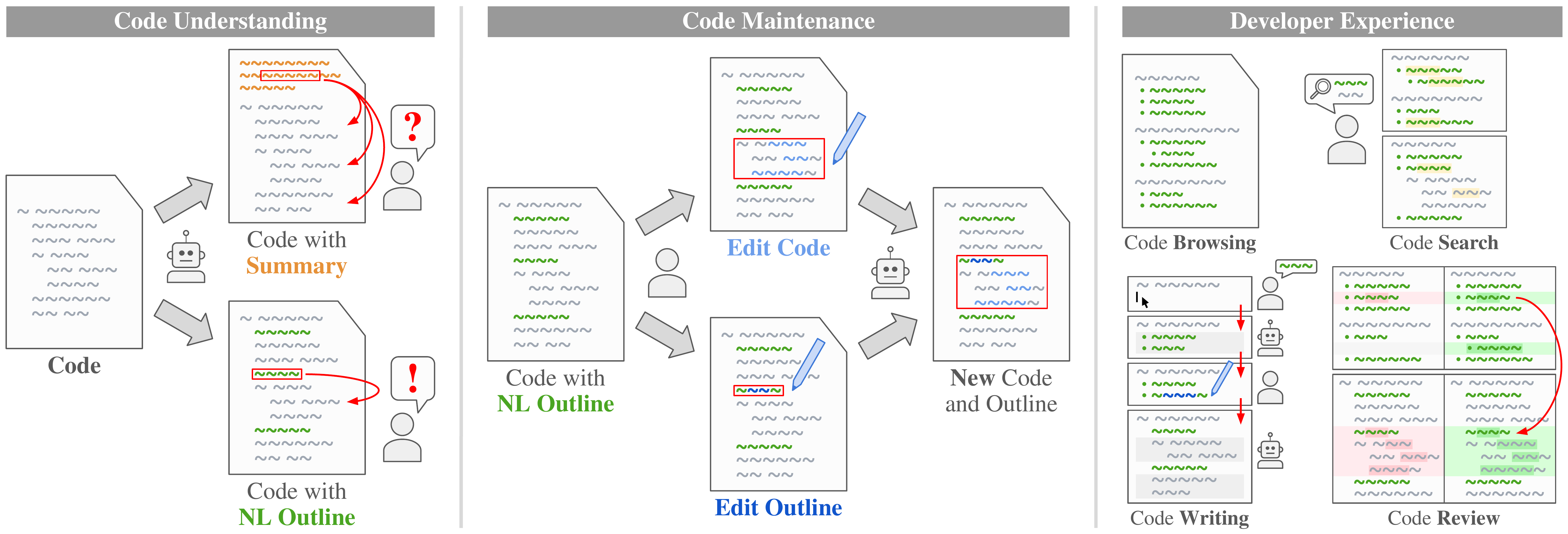}
    \vspace{-6mm}
    \caption{NL outlines can enable a huge variety of AI-based developer assistance features. \enskip\emph{Code Understanding:} An LLM can generate an NL outline for a code function, providing a high-level overview (as in \cref{subfig:interleaved-outline}) that is aligned with the code and more interpretable than a paragraph summary. \enskip\emph{Code Maintenance:} After the developer begins to edit the code \emph{or the outline}, the LLM can complete the edit to keep everything in sync (as in \cref{fig:finish-changes}), offering both automatic documentation and programming via NL. \enskip\emph{Developer Experience:} NL outlines can accelerate code browsing and navigation, enable code search from natural language and contextualize the results, allow users to preview and steer code generation, and summarize changes during code review.}
    \label{fig:diagram}
    \vspace{1mm}
\end{figure*}

\begin{figure*}
    \centering
    \begin{minipage}{.491\textwidth}
    \begin{subfigure}{\textwidth}
      \centering
      \begin{lstlisting}[style=MyPython]
def nearest_neighbor_tour(nodes):
  distances = scipy.spatial.distance_matrix(nodes, nodes)
  current_node = 0
  tour = [current_node]
  tour_cost = 0.0
  distance_to_start = distances[current_node].copy()

  for _ in range(len(nodes) - 1):
    distances[:, current_node] = np.Inf
    neighbor = distances[current_node].argmin()
    tour_cost += distances[current_node][neighbor]
    tour.append(neighbor)
    current_node = neighbor

  tour_cost += distance_to_start[current_node]
  return tour_cost, tour
      \end{lstlisting}
      \caption{One must carefully read the code to understand this function.}
      \label{subfig:plain-code}
    \end{subfigure}
    \par\vspace{5.3mm}
    \begin{subfigure}{\textwidth}
      \centering
      \begin{lstlisting}[style=MyPython,escapechar=!]
nearest_neighbor_tour(nodes):!
    \normalfont\footnotesize\vspace{-2pt}
    \begin{itemize}[leftmargin=2em,labelsep=0.5em,label={\footnotesize\raisebox{0.05em}{$\bullet$}}]
        \item\strut Compute all pairwise distances between nodes.
        \item\strut Initialize the tour.
        \item\strut Iteratively add all nodes to the tour.
        \begin{itemize}[nosep,leftmargin=2em,labelsep=0.5em,label={\footnotesize$\circ$}]
            \item\strut Mark the current node as visited.
            \item\strut Extend the tour by going to the nearest unvisited neighbor.
        \end{itemize}
        \item\strut Complete the cycle back to the starting node.
    \end{itemize}!
      \end{lstlisting}
      \caption{An NL outline for the function, shown without code. The outline concisely conveys the high-level ideas and follows the code structure.}
      \label{subfig:standalone-outline}
    \end{subfigure}
    \end{minipage}\hfill
    \begin{minipage}{.491\textwidth}
    \begin{subfigure}{\textwidth}
    \centering
      \begin{lstlisting}[style=MyPython]
def nearest_neighbor_tour(nodes):
  #* Compute all pairwise distances between nodes.
  distances = scipy.spatial.distance_matrix(nodes, nodes)
  #* Initialize the tour.
  current_node = 0
  tour = [current_node]
  tour_cost = 0.0
  distance_to_start = distances[current_node].copy()

  #* Iteratively add all nodes to the tour.
  for _ in range(len(nodes) - 1):
    #* Mark the current node as visited.
    distances[:, current_node] = np.Inf
    #* Extend the tour by going to the nearest unvisited neighbor.
    neighbor = distances[current_node].argmin()
    tour_cost += distances[current_node][neighbor]
    tour.append(neighbor)
    current_node = neighbor

  #* Complete the cycle back to the starting node.
  tour_cost += distance_to_start[current_node]
  return tour_cost, tour
      \end{lstlisting}
      \caption{An NL outline for the function, shown as comments interleaved with the code, using {\color{starcomment-green}\ttfamily \#*} ``star comment'' syntax (\cref{sec:discussion}). The outline partitions the code into logical sections, provides visual structure, and accelerates code understanding and navigation. Modern LLMs can generate high-quality NL outlines as a form of AI assistance, adding helpful annotations while preserving the code as the source of truth.}
      \label{subfig:interleaved-outline}
    \end{subfigure}
    \end{minipage}
    \vspace{-2mm}
    \caption{An example NL outline for a Python function (a), either displayed without code (b) or interleaved with code (c).}
    \label{fig:nl-outlines}
    \vspace{-2mm}
\end{figure*}

\begin{figure*}
    \centering
    \includegraphics[width=0.91\textwidth,trim={2mm 7.5cm 15cm 16.8cm},clip]{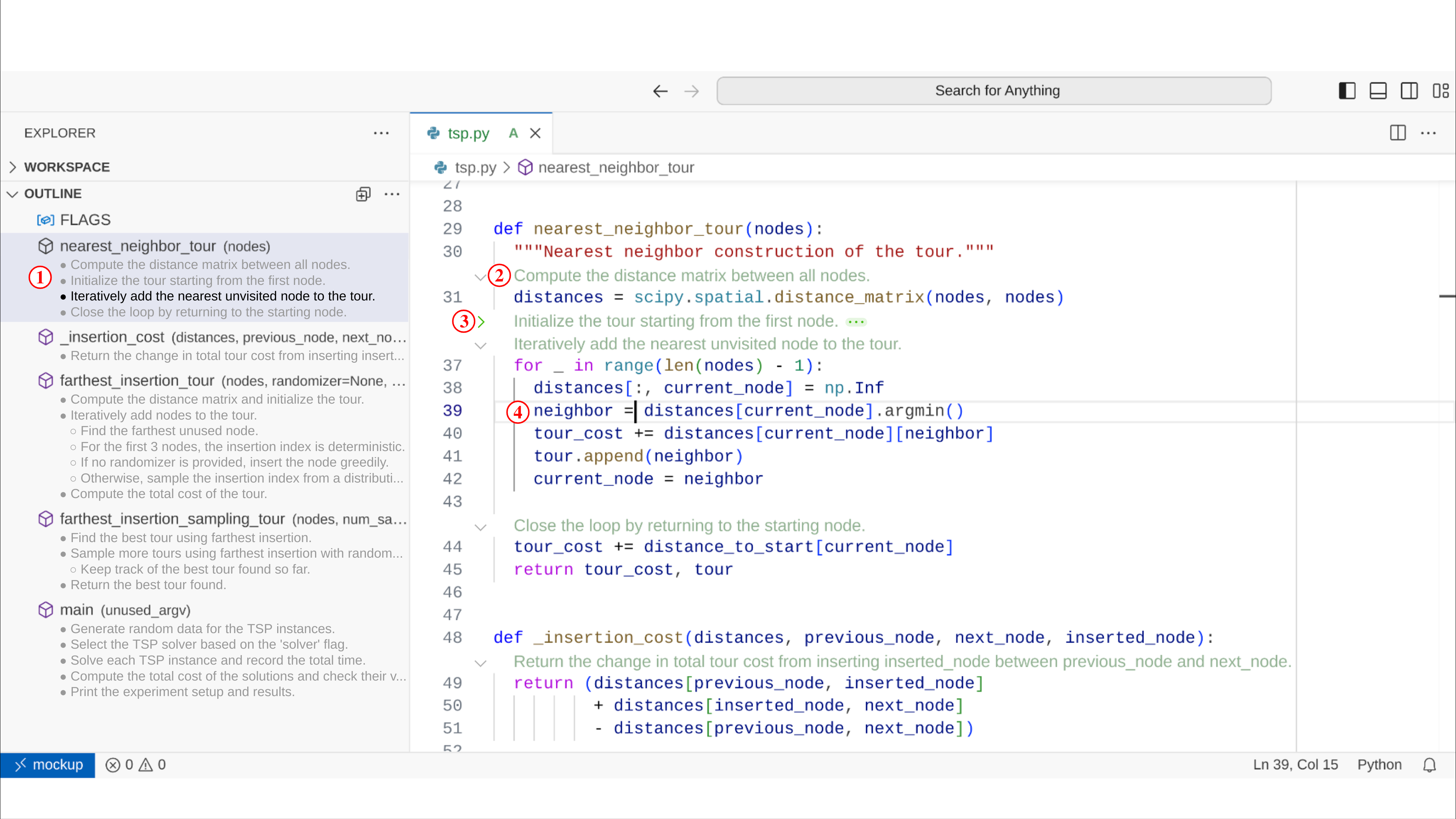}
    \vspace{-2mm}
    \caption{A mockup of how NL outlines could be used in an IDE, with real outlines predicted by Gemini 1.5 Flash. \enskip\circled{1} NL outlines for each function can be displayed in the list of symbols. Clicking an outline statement can move the cursor to the corresponding code location. \enskip\circled{2} In the main editor, NL outlines can be shown interleaved with the code, either as non-code annotations or as code comments. \enskip\circled{3} NL outlines can offer intuitive code folding. \enskip\circled{4} NL outlines can help navigation, e.g., by emphasizing the current outline statement or via shortcuts to jump to the previous or next outline statement.}
    \label{fig:vs-code}
    \vspace{-1mm}
\end{figure*}

Using these themes of outlining, summarization, and literate programming, we propose using LLMs to generate a new form of code explanation: \emph{natural language outlines} (NL outlines). An NL outline consists of natural language statements that partition the code into logical sections, while summarizing each section of code with concise wording. The outline may be shown interleaved with the code or displayed separately, depending on the application. NL outlines are a particularly appealing modality for AI developer assistance due to their broad applicability: in addition to helping developers understand and navigate code, \cref{fig:diagram} illustrates our vision of how NL outlines can be used throughout the software development, e.g., for code generation and maintenance, code search, and code review.

NL outlines are not just one tool or feature. Instead, NL outlines are a conceptual foundation and framework enabling many possible features across all of the different software development surfaces that we use today.
This highlights a key conceptual difference to previous work.
Existing AI auto-commenting techniques focus on reproducing current developer behavior by
training on large code bases. In contrast, our work 
does \emph{not} aim to reproduce existing 
comments, but to encourage the use of a \emph{new style of code summarization}, the NL outline. 
At a technical level, we cannot
simply rely on the LLM's inherent commenting abilities---we need to provide the model with guidance about what makes a good NL outline, which we do via carefully designed few-shot prompting (\cref{sec:generation}).
Thus, this paper aims to demonstrate that NL outlines are \emph{broadly useful} across the software development stack and that \emph{LLMs are already capable enough} to support these new ideas in practice.

\emph{Contributions:} We introduce NL outlines (\cref{sec:outlines}), describe their extensive use cases (\cref{sec:use-cases}), and propose methods for generating them with LLMs (\cref{sec:generation}). We perform a developer study using real code (\cref{sec:experiments}), finding that $60\%$ of outlines were rated by professionals as \emph{excellent} overall ($90\%$ were \emph{acceptable} or better). In two case studies we apply NL outlines toward mobile app security and code review (\cref{sec:case-studies}), finding for example that $26$ of $30$ professional reverse engineers said that outlines are \emph{very} or \emph{extremely} helpful for malware detection, and that $83.8\%$ of our generated outlines were rated as \emph{completely correct}. Finally, we discuss integrating and improving NL outlines in practice (\cref{sec:discussion}).

\section{Natural Language Outlines}
\label{sec:outlines}

A \emph{natural language outline} (NL outline) for a function or snippet of code consists of one or more statements written in concise prose, which are aligned with the code and partition it into semantically-coherent sections, such that each outline statement summarizes the section of code it aligns to.

As an example, \cref{subfig:plain-code} shows a Python function from an author's prior project. Even though the code is self-contained and only uses common libraries, it still takes a considerable amount of time for a developer to parse the code and decipher the steps taken to achieve its goals. A docstring might help somewhat, but docstrings usually discuss the function's contract instead of its implementation strategy which may be more pertinent when editing the code.

To gain a high-level understanding of the implementation, one may instead read the NL outline in \cref{subfig:standalone-outline} which mirrors the code's ideas and structure, while being more concise (using about half as many characters) and easier to read (English versus code). The compact nature of the outline would also reduce scrolling when browsing through the file. If one wishes to see more details or reference the code ``source of truth,'' the outline may also be displayed as comments or annotations interleaved with the code in the style of \emph{literate programming}~\cite{LITERATE}, as shown in \cref{subfig:interleaved-outline}. In this format, the outline provides visual structure and allows the reader to seamlessly switch between reading NL or code syntax as needed. Some developers already write these kinds of comments manually,
even despite the risks that they might become stale and are time-consuming to maintain (but \cref{sec:use-cases} offers solutions for both),
adding to the familiarity of NL outlines as a concept.

But with recent advancements in large language models (LLMs), there is little need for developers to write NL outlines manually, as modern LLMs can generate high-quality NL outlines even for proprietary code outside their training data (\cref{sec:experiments}). Thus, we propose that NL outlines can be used as a means to provide AI \mbox{assistance} to developers, with many use cases illustrated in \cref{fig:diagram} and further discussed in \cref{sec:use-cases}. NL outlines are designed to \emph{improve developer efficiency} for code authors, maintainers, and users.

NL outlines serve a very different purpose than existing automatic code summarization techniques (see \cref{sec:related}).
Compared to generating a \emph{docstring} which describes a function's contract for API users~\cite{PEP257},
an NL outline instead summarizes a function's implementation for developers maintaining or extending it.
Even though NL outlines can be displayed alongside the code like \emph{inline comments} (\cref{subfig:interleaved-outline}), they differ in content and design. Typical inline comments contain sparse notes about design rationales, subtle pitfalls, and remaining tasks~\cite{CLASSIFYING-COMMENTS,TOTT-COMMENTS}, but \citet{UNDERSTANDABILITY} found that such comments are not correlated with perceived or actual understanding of code. In contrast, NL outlines focus on boosting high-level understanding of the code, are overwhelmingly rated as ``Very helpful'' or ``Extremely helpful'' (\cref{fig:outline-quality-survey,fig:security-helpfulness}), and are designed to provide a complete overview even when shown without code (\cref{subfig:standalone-outline}, e.g., in a code search UI).
\cref{app:taxonomy} has further discussion including a taxonomy of different forms of code explanation.

NL outlines are \emph{not} intended to do any of the following: document all details about the code (providing too much information would reduce efficiency), educate developers about concepts or APIs (developers can look up tutorials or documentation elsewhere as needed), provide complex reasoning (some comments are too important or complex to relegate to AI, including documentation of tricky implementation or design details), or replace code entirely (code remains the source of truth considering the ambiguity of NL).

\section{Use Cases for NL Outlines}
\label{sec:use-cases}

The broad applicability of NL outlines is a key factor in their appeal as a surface for human-AI interaction. Here we provide an overview of many potential use cases as illustrated in \cref{fig:diagram}. A full evaluation of each use case is beyond the scope of this paper---instead, \cref{sec:experiments} evaluates the common first step of generating NL outlines to begin with, and \cref{sec:case-studies} explores two concrete applications in practice.

\textbf{Code understanding and navigation.}\quad
\cref{fig:vs-code} is a mockup of how NL outlines could be used in an IDE. NL outlines can be shown in the list of symbols to aid whole-file understanding while being clickable for precise navigation. Outlines can also be shown in the main editor interleaved with the code, assisting with rapid code understanding and offering intuitive code folding~\cite{AUTOFOLDING}. The user experience can be enhanced through customization of outline styling, and IDE shortcuts can show or hide outlines, generate new outlines, convert between outlines and comments, and jump to adjacent outline statements. NL outlines can be similarly displayed in other developer tools for code browsing, searching, and so on.

\begin{figure*}
    \centering
    \begin{subfigure}{\textwidth}
        \centering
        \includegraphics[width=\textwidth,trim={0mm 0.5mm 0mm 0.5mm},clip]{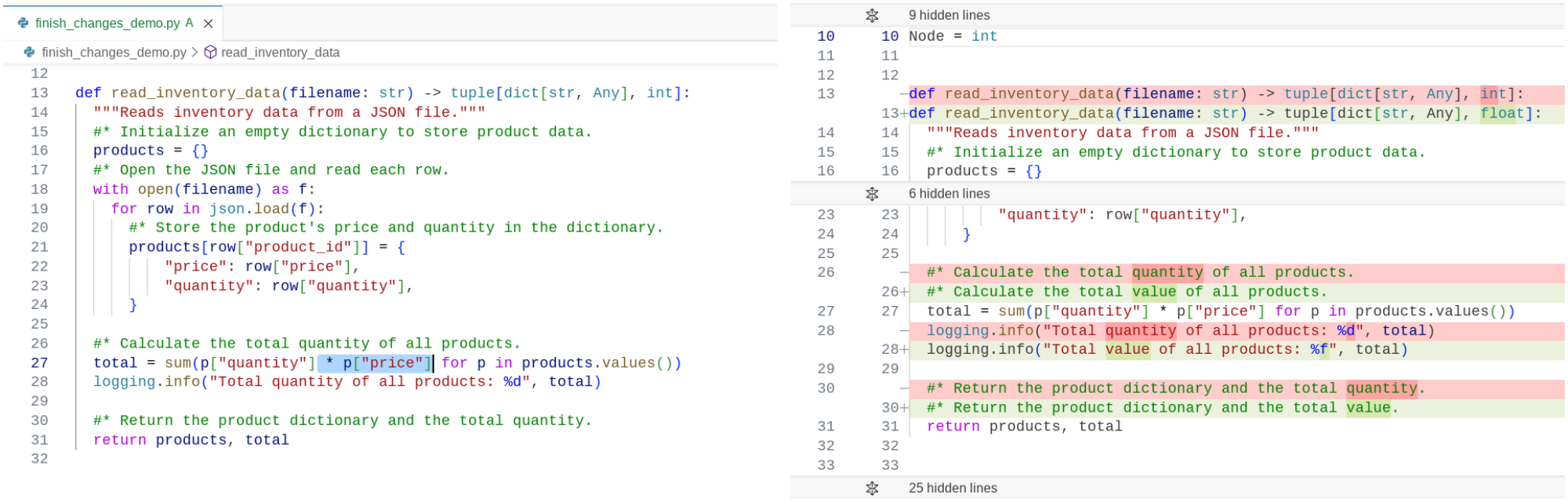}
        \caption{\emph{Left:} A function with an NL outline that correctly describes reading a JSON file and computing a total. The user starts a code change by typing the highlighted text, changing the calculation of \code{total}. \enskip\emph{Right:} The user then invokes the \emph{Finish Changes} feature. The model's prediction is shown as a diff, which correctly updates ``total quantity'' to ``total value'' in the outline and logging statement, and updates the type of \code{total} from \code{int} to \code{float} in the return type annotation and in the logging statement (changing \code{\%d} to \code{\%f}).}
        \label{subfig:finish-changes-a}
    \end{subfigure} \\[8pt]
    \begin{subfigure}{\textwidth}
        \centering
        \includegraphics[width=\textwidth,trim={0mm 7.5mm 0mm 0.5mm},clip]{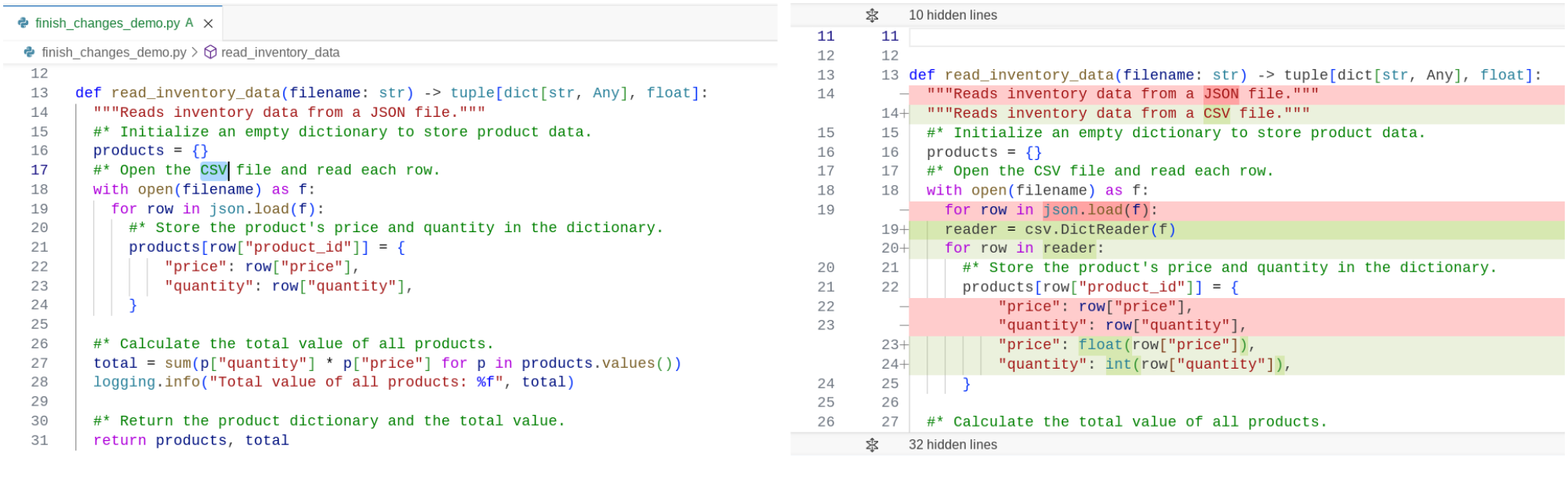}
        \caption{\emph{Left:} After applying the diff in \cref{subfig:finish-changes-a}, the user starts the next edit by changing ``JSON'' to ``CSV'' in the outline as highlighted. \enskip\mbox{\emph{Right:} The} \emph{Finish Changes} feature then provides a diff that correctly updates the docstring and replaces the JSON logic with analogous CSV logic, including adding string-to-number conversions which were not needed before.}
        \label{subfig:finish-changes-b}
    \end{subfigure}
    \vspace{-6mm} 
    \caption{Example usage of a \emph{Finish Changes} prototype feature using Gemini 1.5 Flash. The user can concisely specify the key idea of a change (by editing the code as in \cref{subfig:finish-changes-a} or the outline as in \cref{subfig:finish-changes-b}) and let the LLM finish the job by propagating the changes. \cref{app:finish-changes} contains another more complex example.}
    \label{fig:finish-changes}
    \vspace{-1mm}
\end{figure*}

\textbf{Code maintenance.}\quad
Once the user is accustomed to using NL outlines interleaved with code as in \cref{fig:vs-code}, we can apply NL outlines beyond static (unchanging) code and toward code maintenance.

For example, the IDE can detect when the user is editing a function. When there is a lull in typing, the IDE may automatically update the outline to reflect the current code. (This can be done by providing an LLM with the old code, old outline, and current code, and prompting the LLM to predict an updated outline.)
Some users may prefer that outlines are updated automatically with minimal friction, while others might explicitly request an outline update through a keyboard shortcut and review a diff of the predicted changes. Either way, this use case resolves a common issue of comments becoming stale as the code evolves~\cite{panthaplackel2020learning,wen2019large}, lifting the burden of tedious updates (and constantly checking if any comments need updating) from the user and delegating it to the LLM. Hence, NL outlines can serve as \emph{automatic living documentation} of the code.

We can also perform updates in the reverse direction. If NL outlines are themselves editable by the user, then the user may specify the high-level idea of a code change simply by editing the outline, letting the LLM predict the corresponding code changes.
This concept is quite flexible: edit an outline statement to convey a minor logic change in that portion of the code, add new outline statements to guide a multi-line suggestion of new code, reorder outline statements (potentially outside the function's boundaries) for a refactoring that handles dataflow dependencies better than mere copy/paste, and delete outline statements for a code cleanup that also removes resulting dead code elsewhere. 
This would allow developers to \emph{program primarily through natural language}, as if developing high-level pseudocode that is automatically synced to real code (but we still encourage careful review of the LLM's code).

We can actually generalize the two previous ideas (where the user makes a \emph{complete} edit to the code \emph{or} outline, and the LLM only updates the other to match). Instead, we can let the user \emph{start any change(s)} (to the code or outline) and have the LLM \emph{finish the change(s)} (modifying the code and outline as needed). Many code changes start with one key idea and then require propagating changes to code and documentation elsewhere. With this feature, the user only needs to specify the main idea of each desired change, and the LLM can perform the more tedious work of implementing and propagating the change. This may increase developer efficiency by allowing them to think at a higher level of abstraction and delegating the lesser details (with careful review). The user can edit the outline for broad ideas or edit the code to specify more exact changes, seamlessly switching between code and NL as needed, as in the literate programming paradigm but with a novel bidirectional connection between NL and code powered by modern LLMs.

We built this into a prototype \emph{Finish Changes} feature in an IDE by providing the old code, old outline, current code, and current outline to an LLM which then predicts the new code and new outline. \cref{fig:finish-changes} shows screenshots with real predictions from Gemini 1.5 Flash.
In \cref{subfig:finish-changes-a}, editing a variable assignment causes updates to its description in the outline and its type throughout the code. In \cref{subfig:finish-changes-b}, replacing one word in the outline causes a multi-line implementation change and a docstring update.
Chain-of-thought (CoT) prompting~\cite{COT} is quite effective here, where the LLM lists the user's changes, reasons about further edits to finish those changes, and lastly predicts the new code with new outline interleaved.

\textbf{Code generation.}\quad
The standard approach to function generation is to take a specification (e.g., the function signature and docstring) and predict the entire body, but it can be difficult to understand incorrect or unexpected predictions. The user might even discover that the prediction is so wrong that they delete it and start over, having wasted their time completely.

Instead, as previously mentioned, users can write new outline statements to steer code generation using a \emph{Finish Changes} feature. But when writing an entire new function, we can further assist users by having the LLM \emph{propose} an NL outline (displayed without code) for the user to approve before proceeding---an easier task than writing an outline from scratch or reviewing predicted code directly~\cite{ABSTRACTION-GAP}. The LLM may also predict multiple outlines with different approaches or interpretations for the user to brainstorm with.
The user may edit an outline to correct mistakes instead of debugging the code later. By having the LLM generate code aligned with a user-approved outline, we increase the likelihood that the code satisfies the user, and it is also easier for the user to interpret the code because they already expect the outlined approach.

\textbf{Code review.}\quad
During code review, a key part of software development~\cite{CODE-REVIEW-microsoft,CODE-REVIEW-google}, NL outlines can help reviewers understand code written by others. For example, if NL outlines are automatically updated to reflect code changes, then the outline changes form an NL summary of the code changes. Reviewers may then inspect a diff of the old versus new outlines in addition to the code diff.

\textbf{Code search.}\quad
Code search systems in practice mainly use structured forms of search~\cite{GOOGLE-CODE-SEARCH,CODE-SEARCH}, but users sometimes do not know the needed keyword or regex. NL outlines may enable or enhance more intuitive text-based search:
after generating and indexing NL outlines, a search query containing NL and/or regexes could be matched to outlines and/or code. Users could then search for \emph{concepts}, e.g., to find usage examples or where a particular action occurs in an application. NL outlines can also provide helpful context around search results, whether for NL queries, traditional regex queries, or even symbol references and function call sites.

\section{Outline Generation}
\label{sec:generation}

We can generate NL outlines by few-shot prompting an LLM, but different prompting techniques have different benefits.

\textbf{Interleaved Generation.}\quad
The most straightforward way to generate NL outlines is to provide an LLM with the code to be outlined, prompting it to repeat the code with outline comments added but without any other code changes. This is more effective with prompt instructions describing the purpose of NL outlines and their desirable qualities, and with few-shot examples demonstrating the task and what good NL outlines look like. We call this approach \emph{Interleaved Generation} because the model predicts an outline interleaved with the code. However, a notable downside of this approach is the possibility that the model changes the code against our wishes. If the model's prediction significantly deviates from the original code, the outline might not describe the original code anymore. Minor deviations (e.g., changes to blank lines and existing comments are common) can be ignored if the model gets back on track, but in extreme cases the outline must be discarded.

This issue can be remedied with the \emph{Constrained Generation} approach, using constrained decoding~\cite{FST,GUIDED-GENERATION} to alter token probabilities in a way that prevents changing the code. Specifically, we construct a regex-like constraint that repeats the code exactly, except with optional comment lines between code lines. \cref{app:constrained-generation} discusses additional heuristics and implementation difficulties.

\textbf{Line Number Infilling.}\quad
A drawback shared by Interleaved Generation and Constrained Generation is that they spend output tokens repeating the code, leading to unnecessary cost and latency.
We design a solution called \emph{Line Number Infilling}: we prepend each line of the original code with its line number, and we prompt the model to output a sequence of outline statements, each containing the line number where the outline statement should be added, and the text of the statement itself. For example, the original code \llmtext{def sq(x):\textbackslash n\ return x**2} is given to the model as \llmtext{1|def sq(x):\textbackslash n2|\ return x**2}, and the model may respond with \llmtext{2|Squares the input.} (using multiple lines for multiple outline statements). This prediction task is similar to fill-in-the-middle (FiM) in language modeling~\cite{INFILLING-2020,INFILLING-openai,UL2}, where the LLM must predict text (like outline statements) to insert at specially-marked locations (like the line numbers), although in our case the LLM must also choose which locations to insert at. Since this response format does not involve code tokens at all, we resolve the prior problems of changing the original code or spending tokens to repeat the code. The prompt is also shorter because the few-shot examples do not repeat the code either. In practice, we observe that Line Number Infilling provides a $4\times$ to $5\times$ speedup over Interleaved Generation. However, Line Number Infilling has a downside that the prediction format is less natural for LLMs, i.e., ``code with comments'' is more similar to the training data than ``line numbers and summaries.'' Hence, this approach can cause weaker LLMs to generate outlines with formatting mistakes, misplaced statements, or generally worse quality.

\cref{sec:discussion} has suggestions for further research to improve outline generation, e.g., retrieval-augmented generation~\cite{RAG,RAG-survey,nashid2023retrieval,li2024enhancing}, few-shot selection~\cite{FEW-SHOT-SELECTION,SKILL-BASED-FEW-SHOT-SELECTION}, and finetuning~\cite{RLHF-2017,RLHF-anthropic,RLHF-survey}.

\section{Experiments}
\label{sec:experiments}

We investigated (a) the \emph{rate of formatting issues} in LLM predictions and (b) the \emph{quality of generated outlines} in several dimensions, depending on the LLM and generation technique used. First, we asked 6 professional software engineers and researchers to curate a dataset of 30 Python functions from 21 real projects,\footnote{Five of the dataset contributors are also authors of this paper. There was a bias toward research projects simply due to the contributors' background.} emphasizing variety in libraries used and function kinds (standalone functions, methods, constructors, unit tests, main functions, etc.). Each function had 10 to 90 lines of code (median 46.5 LOC). To mimic realistic usage, we removed all existing outline-like comments\footnote{The LLMs in our experiments were not trained on the removed comments.} (present in 7 functions) but kept other comments including TODOs, developer notes, and comments to disable warnings (in 10 functions).

We tried 5 LLMs: Gemini 1.0 Pro and Ultra~\cite{GEMINI-1.0}, Gemini 1.5 Flash and Pro~\cite{GEMINI-1.5}, and DeepSeek-Coder-Instruct 33B~\cite{DSC}, all with greedy decoding.
We generated outlines using Interleaved Generation and Line Number Infilling,\footnote{We omit Constrained Generation since it is nearly the same as Interleaved Generation for strong models and difficult to set up for older or weaker models.} for a total of 10 outlines per function. We used a fixed set of 8 handwritten few-shot examples.
\cref{app:prompts} contains example prompts.

\begin{table}
\centering
\caption{Parsing Errors and Number of Outline Statements}
\label{tab:outline-stats}
\vspace{-1mm}
\begin{tabular}{@{}l@{\hspace{\tabcolsep}}l | c@{\hspace{\tabcolsep}}c@{\hspace{\tabcolsep}}c | c@{}}
\toprule
Technique                & Model                       & None & Minor & Major & Avg Stmts \\
\midrule
\multirow{5}{4.5em}{Interleaved Generation} & Gemini 1.5 Pro             &        \textbf{28} &            \phantom{0}2 &            0 & 4.20 \\
 & Gemini 1.5 Flash                &        27 &            \phantom{0}3 &            0 & 5.63 \\
 & Gemini 1.0 Ultra                &        20 &            \phantom{0}7 &            3 & 4.90 \\
 & Gemini 1.0 Pro                 &        15 &           12 &            3 & 7.63 \\
 & DeepSeek-Coder &        16 &           10 &            4 & 5.70 \\
\midrule
\multirow{5}{4.5em}{Line Number Infilling}    & Gemini 1.5 Pro             &        \textbf{30} &            \phantom{0}0 &            0 & 4.23 \\
 & Gemini 1.5 Flash                &        \textbf{30} &            \phantom{0}0 &            0 & 4.30 \\
 & Gemini 1.0 Ultra                &        \textbf{30} &            \phantom{0}0 &            0 & 4.70 \\
 & Gemini 1.0 Pro                 &        \textbf{30} &            \phantom{0}0 &            0 & 7.00 \\
 & DeepSeek-Coder &        24 &            \phantom{0}5 &            1 & 5.97 \\
\bottomrule
\end{tabular}
\vspace{-1mm}
\end{table}

Because each LLM response is just text, we must programmatically parse the text to extract the predicted outline. During parsing, we may encounter errors in the prediction's formatting (e.g., changing the code in Interleaved Generation) or other problems with the outline. We classify these parsing errors into major errors that affect outline quality and minor errors that we can recover from without impacting outline quality.\footnote{\cref{app:errors} lists the specific errors we checked for.}
\cref{tab:outline-stats} provides the \emph{rate of formatting issues} in terms of functions out of 30 for which the prediction had no parsing errors, only minor errors, or a major error. We observe that the Gemini 1.5 models consistently produce well-formatted predictions (with zero major errors) while the other LLMs make more mistakes. \cref{tab:outline-stats} also shows the average number of statements in the predicted outlines. Too many statements often means poor abstraction of ideas, but too few may indicate a lack of detail.

\begin{figure}
    \centering
    \includegraphics[width=\columnwidth,trim={2mm 2mm 2mm 2mm},clip]{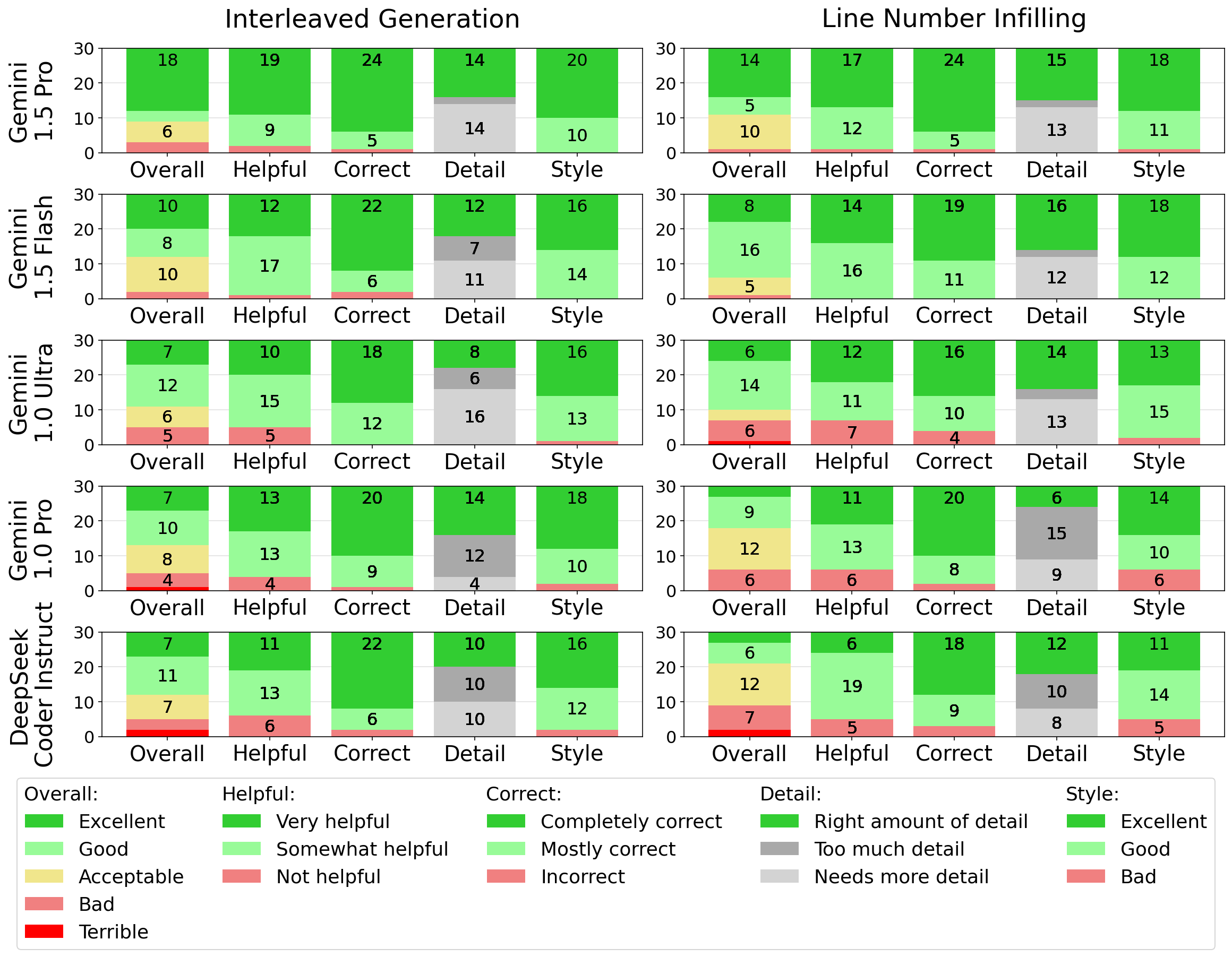}
    \caption{Quality survey results, showing the frequency of answers to survey questions for outlines produced by a given LLM and generation technique.}
    \label{fig:outline-quality-survey}
\end{figure}

We performed surveys about the \emph{quality of generated outlines}, presenting each function's 10 outlines in shuffled order to the function's contributor, i.e., someone very qualified to rate the outlines because they deeply understand the function and its purpose. For each outline, we asked about overall quality, helpfulness, correctness, amount of detail, and style/fluency;\footnote{\cref{app:survey-questions} lists the survey questions. \cref{app:example-outlines} contains several examples of high-quality and low-quality outlines and their survey responses, with commentary.} the results are in \cref{fig:outline-quality-survey}.

Even though the vast majority of dataset functions refer to custom dependencies (e.g., helper functions) which are not in the prompt or training data, we find that LLMs are nevertheless very capable of generating NL outlines that are high-quality, helpful, and correct. For example, using Gemini 1.5 Pro and Interleaved Generation, $60\%$ of outlines were rated as having \emph{excellent} overall quality ($90\%$ were \emph{acceptable} or better), $63\%$ were \emph{very helpful}, and $80\%$ were \emph{completely correct}. Some outlines had minor issues (e.g., $17\%$ were \emph{mostly correct}), but poor performance was quite rare: $3$ of $30$ outlines were \emph{bad} quality, $2$ were \emph{not helpful}, and only $1$ was \emph{incorrect}. The Line Number Infilling technique led to slightly worse predictions (but it is several times faster, see \cref{sec:generation}). Gemini 1.5 Flash was the next best LLM and the others were noticeably worse.

Interestingly, the amount of detail depends heavily on the LLM used. We initially iterated on prompts using Gemini 1.0 Pro, which we found often includes too much detail (as reflected in the surveys). We addressed this with prompt instructions and by reducing detail in the examples, which anecdotally had some intended effect for 1.0 Pro but causes 1.5 Pro to provide too little detail (as 1.5 Pro seems much better at following the prompt). Further LLM-specific prompt engineering would likely improve the level of detail.

We emphasize that this evaluation uses real functions from real work. Overall, the results suggest that state-of-the-art LLMs are strong enough to generate helpful NL outlines in practice.

\section{Case Studies}
\label{sec:case-studies}

We present deep dives into two concrete applications.

\subsection{Android Security}
\label{subsec:android-security}

We consider the problem of assessing the security and privacy of Android applications (or apps). These assessments are performed routinely by online app stores such as Google Play and Samsung Galaxy Store, where apps submitted by developers across the world are analyzed for compliance with
policies\footnote{Google Play's policies are at \url{https://play.google/developer-content-policy/}.}
meant to protect app users, e.g.:
\emph{
    Apps that are deceptive, malicious, or intended to abuse or misuse any network, device, or personal data are strictly prohibited.
}

Determining whether an app is deceptive or malicious involves understanding the code and relating its functionality to the app's description, UI, and user expectations.
There are millions of apps in stores and tens of thousands of new apps added per month, so security assessments must be efficient.

To detect malicious apps, reverse engineers (REs) use static and dynamic program analyses to flag code as potentially problematic and then confirm violations through manual review. The initial challenge is in the volume of code: many apps have millions of lines of code, thousands of classes, and complex event-driven execution paths. Furthermore, an app's source code is usually not available, so REs inspect \emph{decompiled code} which is difficult to understand for several reasons. Decompiled app code is unnatural due to the compilation process performing optimizations (e.g., inlining code and changing code structure) and removing code artifacts (e.g., comments, docstrings, types, variable names, and names of inlined API constants). App code may be further transformed by minification to reduce app size~\cite{PROGUARD} and obfuscation to protect intellectual property~\cite{DEXGUARD}. Nevertheless, REs must quickly understand the code to find policy violations---a promising application for NL outlines.

\textbf{Applying NL outlines.}\quad
We collected a dataset of 80 decompiled functions from real Android apps. Half were manually curated ``suspicious'' functions that REs had previously flagged as using malware techniques, and we focused on including a variety of techniques and avoiding similar code. The other half were ``benign'' functions randomly sampled from 8 popular apps assumed to be malware-free, all from different companies and genres. All dataset functions have between 24 and 99 lines of code, and we tried to make the two halves of the dataset have similar distributions in code length and style (minified or obfuscated) as much as reasonably possible to reduce bias.

We then used Gemini 1.0 Ultra to generate 3 kinds of predictions per function, designed for different benefits. \mbox{(1) A} \emph{suspicion score} is an integer from 0 (not suspicious at all) to 3 (very suspicious) which is extremely quick to read, allows more nuance than a simple binary prediction,\footnote{The severity of a suspicious function often depends on how it interacts with other functions and the app's expected functionality.} and enables programmatic processing due to its simplicity. \mbox{(2) A} paragraph-form \emph{summary} of about 1-3 sentences describes what the function does and why it is suspicious or not, in a format that is a baseline for LLM-based code explanation and is  familiar to users. \mbox{(3) Lastly}, we generate an \emph{NL outline}, shown with inline comments, which focuses only on the suspicious parts and which is empty by design if the LLM predicts that the function is not suspicious at all. The purpose is to point directly to the problematic areas of code for faster understanding and localization of the malware implementation, without non-security-related distractions (REs usually do not care about the behavior of benign functions).

\newcommand{\boxhighlight}[1]{\adjustbox{varwidth=\columnwidth-15ex,margin=0.5ex,bgcolor=yellow!20}{#1}}
\begin{figure}
    \small \raggedright
    \textbf{Suspicion score:} \highlight{2 / 3}
    \vspace{0.8mm}

    \textbf{Summary:}
    \boxhighlight{This function reads the user's clipboard data. This is suspicious behavior that may be a privacy concern.}
    \vspace{0.8mm}

    Code with \textbf{NL outline:}
    \vspace{1mm}
    \begin{lstlisting}[style=MyJava,escapechar=?]
public void a(android.content.Context p5) {
  ?\highlight{// Accesses the user\textquotesingle s clipboard.}?
  android.content.ClipboardManager v0_2 =
      ((android.content.ClipboardManager)
       p5.getSystemService("clipboard"));
  if (v0_2.hasPrimaryClip()) {
    android.content.ClipData v1_3 = v0_2.getPrimaryClip();
    if (v1_3.getItemCount() != 0) {
      Maybe([Type(UNKNOWN), Primitive(INT)]) v2_3 = 0;
      while (v2_3 < v1_3.getItemCount()) {
        ?\highlight{// Sends the clipboard contents to \`{}this.ab1.c23()\`{}.}?
        this.ab1.c23(v1_3.getItemAt(v2_3).getText().toString());
        v2_3++;
  } } } return; }
    \end{lstlisting}
    \caption{Real LLM predictions in 3 styles (suspicion score, summary, and NL outline) for a decompiled function.
    }
    \label{fig:security-example}
    \vspace{-1mm}
\end{figure}

We obtain all three predictions in a single LLM query with 7 few-shot examples.
The LLM first predicts the summary to help it reason about the other predictions, similar to CoT prompting. The suspicion score is predicted next. The outline is predicted last to make it consistent with the score, i.e., it is consistently empty if and only if the score is 0. \cref{fig:security-example} shows actual LLM predictions for a decompiled function of our creation.

\textbf{Evaluation.}\quad
We aimed to evaluate the \emph{accuracy} and \emph{perceived helpfulness} of each kind of prediction. We first consider the accuracy of predicted suspicion scores.
Of the 40 suspicious functions, four had a predicted suspicion score of 0, which we treat as false negatives, and of the 40 benign functions only two were false positives with nonzero suspicion score (both scores of 1, the smallest nonzero amount of suspicion). Further inspection of these cases led to interesting observations. The two false positives were actually reasonable predictions that correctly describe privacy-impacting behavior in fair terms without raising more alarm than necessary.\footnote{For example, one such summary was: ``This function attempts to get the user's location and determine if it is night time. This is not problematic itself, unless code elsewhere additionally exfiltrates the data.''} In our survey of REs (described below), for both functions, the three REs assessing the function unanimously agreed that the model's predicted suspicion score (of 1) was accurate, and the median of the RE's suspicion scores was also 1, so these cases should not be considered false positives after all. Among the four false negatives, two cases had all three REs agreeing that the predictions were accurate. For example, an RE originally flagged one of those functions as suspicious because it called \code{ActivityManager.getRunningAppProcesses} which can be a privacy concern. However, the LLM used the surrounding code context to correctly deduce that it only obtains the name of the current process and nothing else, which is not actually problematic. This leaves two actual false negatives remaining,\footnote{In one case, the LLM balked at analyzing a severely obfuscated function, but when sampling again, the LLM often predicts that it \emph{is} suspicious due to the obfuscation. Further prompt engineering would improve consistency here.} for a false negative rate of $5\%$ and false positive rate of $0\%$.

To judge the accuracy of the other kinds of predictions, we surveyed 30 professional REs whose primary role is to review Android apps. We split the REs into 3 groups of 10 by their amount of experience. For each function, the LLM's predictions were rated by one RE from each group, so that each RE rated predictions for 4 suspicious and 4 benign functions.\footnote{For each experience group, we shuffled the assignments of functions to REs while balancing the predicted suspicion scores for diversity in predictions.} Prior to seeing the LLM predictions for a function, REs also provided their own suspicion score. Afterward, each RE completed a final survey about the predictions overall.

In the per-function surveys, $95.4\%$ said the LLM's suspicion score was \emph{accurate}, $82.1\%$ said the summary was \emph{completely correct} (versus $17.9\%$ \emph{mostly correct} and $0\%$ \emph{incorrect}), and excluding empty outlines for functions predicted as benign, $83.8\%$ said the outline was \emph{completely correct} ($16.2\%$ \emph{mostly correct}, $0\%$ \emph{incorrect}). About $77\%$ of both summaries and outlines were rated as \emph{right amount of detail} (about $23\%$ \emph{needs more detail}, less than $1\%$ \emph{too much detail}).

We also measured correlations between suspicion scores provided by the LLM and each group of REs. Compared with the scores from the most-experienced REs, the LLM's scores have a Pearson correlation coefficient $r=0.85$, higher than that of the other two groups of REs ($r=0.80$, $r=0.78$). Alternatively, compared with the average of RE's scores, the LLM's scores have $r=0.96$ which is higher than that of any group of REs ($0.92 \le r \le 0.93$). Either way, the LLM performs better than most REs on this task and dataset. Overall, we conclude that \emph{the LLM is very accurate}.

\begin{figure}
    \centering
    \includegraphics[width=\columnwidth]{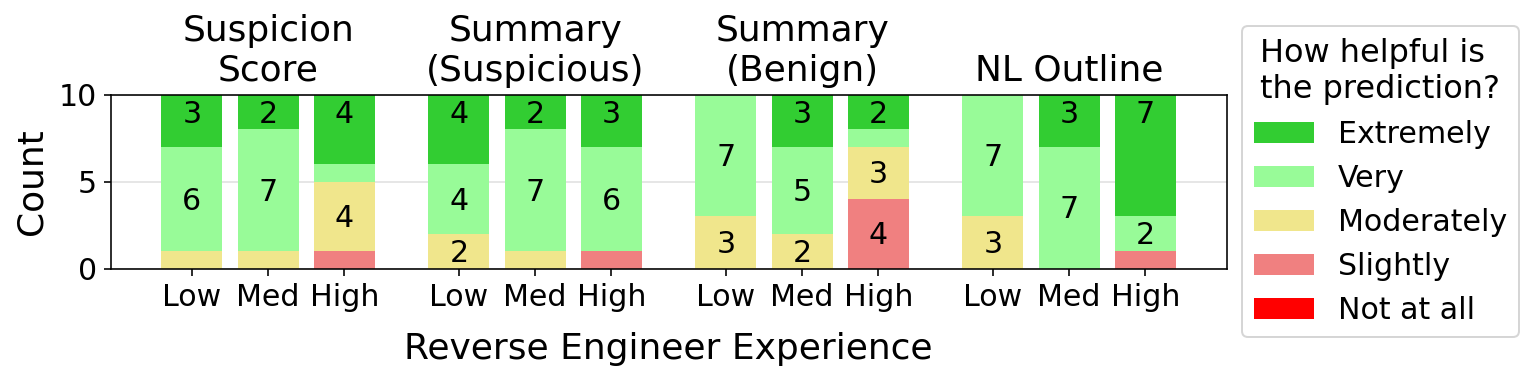}
    \vspace{-4mm}
    \caption{Helpfulness of LLM-predicted suspicion score, summaries (for suspicious and benign functions), and NL outlines (for suspicious functions only). REs were split into three groups of 10 based on their level of experience.}
    \label{fig:security-helpfulness}
    \vspace{-1mm}
\end{figure}

Next, we analyzed the \emph{helpfulness} of each style of prediction in \cref{fig:security-helpfulness}. A majority of REs found every kind of LLM prediction \emph{very} or \emph{extremely} helpful, but some patterns stand out. First, we expected summaries to be more helpful for suspicious functions than benign functions, because a function's behavior is less important if it is not security-impacting. The survey responses (especially from the most experienced REs) confirm our expectation. This also validates our design decision to have NL outlines only mention security-related aspects of functions predicted to be suspicious. Another particularly interesting pattern is that the most experienced REs found the NL outlines to be much more helpful ($70\%$ \emph{extremely} helpful) than the other REs did. This may reflect a difference in RE responsibilities: the most experienced REs are often final decision-makers who need to fully understand a suspicious implementation, and while suspicion scores and summaries can help prioritize focus, only outlines help in understanding the implementation strategy.

REs indicated that summaries and outlines were helpful in all the ways we asked about, as summed up by one RE's feedback---\emph{``As proposed, the LLM help will change the game for us RE's: in speed, accuracy, and confidence.''} Summaries helped in terms of \emph{speed} (as indicated by $100\%$ of REs), \emph{understanding} ($83\%$), and \emph{confidence} ($80\%$), although for the most experienced group of REs, summaries were not as helpful in understanding ($60\%$) and confidence ($50\%$). On the other hand, outlines helped all groups of REs relatively equally, in \emph{speed} ($93\%$), \emph{understanding} ($90\%$), and \emph{confidence} ($80\%$).

Finally, $97\%$ of REs said they would read \emph{both} summaries and outlines if provided (i.e., they are complementary), but one RE in the most experienced group would read \emph{only outlines} and provided the following feedback: \emph{``Score and comments are by far the best features. Summaries would allow quicker reviews, but it is necessary to have a look in the code, where the comments would help a lot in general.''}\footnote{The surveys referred to outlines as ``comments'' as a more familiar term.} It is clear that the NL outlines concept \emph{adds significant value} in this application.

Given the overwhelming success of these experiments, we have started upgrading RE tools to surface the LLM predictions.

\subsection{Code Review}
\label{subsec:code-review}

This application is motivated by a common pain-point in code review: the
difficulty in reviewing large or complex change lists (CLs). With increasing CL complexity, it becomes increasingly difficult to maintain a complete mental model of all of the changes and how they interact. This software development anti-pattern is so substantial that code reviewers may even send large CLs back to the authors to be split into smaller CLs that are individually more manageable. Reviewers actually spend much less time per line reviewing large CLs, raising potential concerns about fatigue and review quality. Although tools have been developed to assist in moving code changes between CLs, actually splitting a CL is time-consuming, and in some cases it might not be possible if the changes are particularly entangled. So, the burden often falls on the reviewer to mentally disentangle messy CLs during code review.

We address this with a prototype feature called \emph{Virtual CL Split}, which is a layer in an existing code review interface that allows a reviewer to view individual code changes grouped into logical topics. Each topic is described with a natural language phrase that helps the reviewer interpret the changes. The topics form a partition of the original CL, but this splitting is done \emph{virtually}, i.e., the CL itself is unchanged and there is no requirement that any individual topic can become a standalone CL. Upon selecting a topic in the UI, all changes in that topic have their code diffs expanded and other changes are collapsed (or shown with muted diff highlighting when the reviewer wants to see more file context), so that the reviewer can focus on the selected topic to simplify their mental stack. Virtual Split ensures that every code change appears in some topic, so that the reviewer is ``done'' after examining all topics.
\cref{app:virtual-split} shows screenshots of the Virtual Split feature on an example CL.

For example, consider a CL where the author first performed a refactoring causing multiple small changes. This enabled the author to implement a complex new function and add unit tests for it. Finally, the author invoked a code formatting tool causing whitespace changes.
In this CL, a Virtual CL Split may suggest 3 topics: first the new function with its tests, then the refactoring, and lastly the whitespace changes. The first topic enables the reviewer to inspect the new function and its tests \emph{together} (without unnecessary searching or scrolling even though they come from different files), and the topic description summarizes the new functionality. This is the most important part of the CL, so it is shown first to improve review quality~\cite{FILE-ORDER}. Afterward, the second topic helps the reviewer inspect the refactoring without distraction from whitespace changes, which are in the last and least important topic.

To generate a Virtual Split, we first provide the CL description and all file diffs (in unified format) to an LLM to generate a list of topics with few-shot prompting. Then, for each file in parallel, we provide the CL description, the diff with line numbers prepended, and the list of topics. With few-shot prompting, the LLM generates an NL outline of the diff using the Line Number Infilling technique, except that the NL outline partitions and describes code changes (instead of code functionality), and for each diff section, the model also predicts which topic it belongs to. The resulting diff section boundaries and topic assignments thus form the Virtual Split. We highlight that this is an example of how NL outlines can serve as a foundation for features with very different objectives and UX.

\textbf{Evaluation.}\quad
In a preliminary market fit study, we worked with four software engineers from the same team to evaluate how often CLs are \emph{complex enough} to benefit from virtual splitting, and how often the predicted splits were \emph{useful}. We randomly selected 73 CLs submitted in their team over the prior weeks with more than 10 lines changed. Many of these were too simple for virtual splitting to be helpful, including CLs that were still very short, CLs which cannot reasonably be split further, and CLs with mostly deletions, generated files, or rollbacks. Overall, $24$ ($33\%$) of the CLs were deemed \emph{complex enough} for a virtual split to be helpful. At the scale of a large software company, this is a huge opportunity to help reviewers facing complex CLs.
We generated virtual splits, and among the $24$ complex-enough CLs, $14$ splits ($58\%$) were described as \emph{useful} during review.
We also observed the pattern that, if the CL is complex enough and the split was rated as \emph{high-quality} (i.e., changes were correctly assigned to topics), then the split was almost always considered \emph{useful}, indicating that the concept is well-designed.

After the preliminary study, we improved split quality via prompting improvements, e.g., summarizing each diff as chain-of-thought reasoning
before generating topics,
better utilizing Gemini's long context,
providing ``hints'' of line numbers where blocks of changes start, and adjusting few-shot examples and instructions. In a small experiment on 10 CLs, these changes noticeably improved split quality for 7 CLs and only decreased quality for 1.
We also used a newer LLM and made other UX and infrastructure improvements, and launched the feature to a beta group of about 7000 users
on live code review traffic.
Examples of positive user feedback include:
\newenvironment{myquote}{\list{}{\leftmargin=1.5em\rightmargin=1.5em}\item[]}{\endlist}
\begin{myquote}
    \emph{``Is this a new feature? It seems to parse the CL description, breaks it down into small individual tasks/features and then only shows the relevant diffs. This is amazing :mind-blown:''}
\end{myquote}\vspace{1pt}
\begin{myquote}
    \emph{``Because the CL was split into 3 tasks, it was easier for me to understand the change step by step.''}
\end{myquote}\vspace{1pt}
\begin{myquote}
    \emph{``It makes it a lot easier to review related changes across multiple files. Without this feature, I have to go back and forth between files a lot more especially for large CLs.''}
\end{myquote}\vspace{1pt}
\begin{myquote}
    \emph{``I saw a CL getting split up into topics in Critique. Simply beautiful... It's already saved me time in code reviews :)''}
\end{myquote}\vspace{3pt}
Negative feedback mainly expressed that the CL was already simple enough to be easily understood without a split, so we are working to better detect these cases preemptively.

\section{Discussion}
\label{sec:discussion}

We discuss practical considerations that arise when integrating NL outlines at scale, ways to improve outline generation, using outlines within LLM pipelines, and limitations.

\textbf{Verifying outlines.}\quad
When paired with feedback mechanisms, NL outlines can become a resource shared across developers. For example, thumbs up/down feedback can be collected, and with sufficient thumbs down, an outline may be hidden or regenerated. Enough thumbs up can cause an outline to become ``verified'' and displayed in a special way (e.g., with a checkmark) signaling that this outline has approval from other humans. Trusted developers such as code authors, code reviewers, and project leads may be given outsized power to influence outlines, e.g., unilaterally marking an outline as verified. Trusted developers may also be allowed to edit outlines to correct mistakes (with the edits shown to everyone else), increasing the quality of outlines and users' confidence in them.

However, verifying NL outlines takes time and effort, which must be considered alongside the quality and confidence benefits of having verified outlines. We emphasize a philosophy of making things easier for developers in the long run. NL outlines as a concept may be more or less helpful for different teams, depending on what they do (e.g., writing new code versus maintaining existing code, or building standalone projects versus frameworks with users), the technology and concepts used (e.g., custom APIs that confuse the LLM), and simply personal preference of team members. A team may reasonably decide that reviewing NL outlines should be completely optional. Ultimately, developers should not be given more work that they think is not worthwhile. We also emphasize that the code is the source of truth: it is what the computer runs, and NL is inherently ambiguous. Further research is needed to measure the perceived and actual benefits of NL outlines used at scale in the various ways proposed, which should then guide decisions around tooling, UX design, and team culture.

\textbf{Comments or metadata?}\quad
When building NL outlines into prototype IDE features, we encountered a major implementation decision: should outlines be stored in the source code as comments with a special syntax, or as non-code metadata that is stored and manipulated separately from code? We discuss these options from the perspective of a large software company.

Storing outlines as non-code metadata has drawbacks mainly in the difficulty of implementation: the metadata files or databases must be managed, each developer tool would need to load or edit the metadata in addition to source code and have a custom UI layer to support displaying or editing outlines, and implementation differences may reduce portability and compatibility.

Alternatively, if NL outlines are to be inserted into code as comments, then they should use a special syntax to distinguish them from other comments, so that readers can treat AI-generated comments with appropriate caution. We propose the \emph{star comment} syntax as in \cref{subfig:interleaved-outline}. This is simply a regular line comment but with an asterisk after the comment syntax, e.g., \commentcode{\#*} or \commentcode{//*} for most common programming languages.\footnote{A star comment \commentcode{\footnotesize //*} should not be confused with a multi-line comment \commentcode{\footnotesize /* ... */}. Syntax highlighting can help differentiate the two.} The asterisk is chosen to mimic the stars commonly found in AI-related icons such as for Gemini and GPT-4, while being ASCII and standard on keyboards (hence compatible with code). \emph{Verified} outlines may be shown with syntax like \commentcode{\#*!} indicating that the comment is more trustworthy. Programmers and developer tools should follow two basic rules about star comments: (1) Star comments are usually AI-generated and should be trusted cautiously, and (2) It is expected that star comments may be changed, added, or removed by AI tools.

Adding NL outlines into code as star comments leads to a relatively simple integration of NL outlines into developer tooling. There is no need to update programming language parsers because star comments are already comments, and similarly no need for metadata outside the source code file. By default, IDEs and other developer tools would display star comments in the same way as regular comments except distinguished by the asterisk, but it would also be simple to adjust their syntax highlighting. Star comments are maximally portable because the syntax is not specific to any developer tool or implementation. Users can add, remove, or edit star comments in the same way as editing other code. Users can also convert between normal comments and star comments by adding or deleting the star to allow or prevent AI updates to that comment. IDEs may offer keyboard shortcuts to perform some of these actions. However, star comments have important downsides in churn, flexibility, and distraction, discussed in more detail in \cref{app:star-comments-downsides}.

\textbf{Improving outlines.}\quad
In this work, we generate NL outlines of a function $f$ only from $f$'s code and few-shot examples, but providing more context through retrieval-augmented generation (RAG)~\cite{RAG,RAG-survey,nashid2023retrieval} may improve the predicted outline's quality. For example, one may provide the LLM with the file containing $f$, unit tests for $f$, functions that call or are called by $f$, and even design documents describing the need for $f$. Similarly, few-shot selection~\cite{FEW-SHOT-SELECTION,SKILL-BASED-FEW-SHOT-SELECTION} can choose few-shot examples that are the most similar to $f$, e.g., considering the programming language, kind of function (unit tests, instance methods, etc.), and so on.

Finetuning is another technique to improve outline generation quality, using data collected through various means. One can mine outline-like comments from existing code, but many such comments are low-quality or incomplete (i.e., only part of the function is outlined), and we found it difficult to detect such cases at scale. Another approach is to sample outlines and keep those that score well on LLM-based evaluation metrics~\cite{LLM-COMPARATOR,RTC}, but this approach is limited by the model's ability to generate a good outline in the first place. A promising direction could be to use real feedback gathered from NL outlines integrated into developer tooling, such as thumbs up or down, accepted or rejected outline predictions, and user edits to outlines. These would enable the outline-generation system to learn and improve from human feedback (RLHF)~\cite{RLHF-2017,RLHF-anthropic,RLHF-survey}.

\textbf{Helping LLMs.}\quad
\cref{sec:experiments,subsec:android-security} show that NL outlines are very helpful to humans---can they also help LLMs? Indeed, NL outlines as an intermediate step can potentially improve LLM performance on downstream code-related tasks. For example, the case study in \cref{subsec:code-review} generates outlines to summarize code changes in individual files, so that a subsequent prompt can reason about a multi-file change without being overwhelmed. Similarly, \cref{app:generalization} describes how one might use function outlines within a larger system to generate NL outlines for files and projects.

\textbf{Limitations.}\quad
In \cref{sec:experiments} we used 6 survey participants, and in \cref{subsec:code-review} we used 4 participants from one team, so those findings may be specific to those participants. As described in \cref{sec:experiments}, the 30 dataset functions were not sampled uniformly. In practice, many functions are boilerplate, very short or simple, or otherwise less interesting, making it less likely that a developer would read an NL outline for that function anyway. Instead, we focused on dataset \emph{variety} and choosing \emph{interesting} functions (and only in Python which is our expertise). Our choice of functions can introduce bias.

\section{Related Work}
\label{sec:related}

Code summarization has attracted much recent attention~\cite{CODE-SUMMARIZATION-survey-2022,CODE-SUMMARIZATION-survey-2019,hu2022practitioners,CODE-SUMMARIZATION-2022,CODE-SUMMARIZATION-2020,ahmed2024automatic,DOCUMENTATION-GENERATION-2024,MULTI-INTENT-COMMENT,CODE-SUMMARIZATION-2024,mu2023developer,COMMENT-CATEGORIES}. \cref{app:taxonomy} presents a taxonomy contrasting the many forms of code explanation. NL outlines are one such form, but instead of a standard paragraph, an outline is written as a list of statements that can be aligned to and interleaved with the code. Moreover, most methods of code summarization have not been applied to the many use cases of NL outlines (\cref{sec:use-cases}).

Some related works do consider individual use cases, e.g., Panthaplackel et al.~\cite{panthaplackel2020learning} aim to update comments given code changes using a custom bidirectional GRU, but NL outlines accomplish this through LLM prompting and generalize to the reverse direction of updating code given outline changes. Liu et al.~\cite{ABSTRACTION-GAP} present a technique for code generation from NL where the generated code is summarized to assist the user in writing future prompts, but they do not consider applications beyond code generation. Code folding~\cite{AUTOFOLDING} can provide a quick overview of code, and NL outlines extend this idea by providing intuitive folding locations and NL summaries of the folded code. A key contribution of NL outlines is their general applicability toward all of these use cases and more.

LLMs have been broadly applied to software engineering~\cite{LLM-SE-survey,LLM-SE-review,LLM-SE-survey-2,IDE-AI-LIT-REVIEW}, including integrating LLMs into IDEs to provide coding suggestions~\cite{COPILOT,XCODE,EXTRACT-METHOD}, chat assistance~\cite{PROGRAMMERS-ASSISTANT,MIXED-REALITY}, educational information~\cite{GILT,EBOOK}, and novel coding experiences~\cite{SPELLBURST,PROMPT-SAPPER}. NL outlines offer a different approach to incorporating AI into developer tools.

\section{Conclusion}
\label{sec:conclusion}
\emph{Natural language outlines} are a new form of code explanation that summarize the main ideas for rapid understanding. Our experiments show that LLMs can generate accurate and helpful NL outlines in practice. We describe the huge potential to use NL outlines to assist software developers in numerous ways, and we call for further research to advance these transformational ideas.

\section*{Acknowledgments}

We thank the following people (listed alphabetically by last name) for various helpful conversations, feedback, assistance, and support: Martin Abadi, Miltos Allamanis, Brian Bosak, Chandler Carruth, Eunice Chang, Nina Chen, Phil Chung, Evan Conaway, Patricia Diaz, Elvina Fan, Felix Gimeno, Chenjie Gu, Rishab Gupta, Juliana Hage Chaim do Val, Matthew Fowles Kulukundis, Jett Lewis, Lihao Liang, Michael Xieyang Liu, Andrew Macvean, Jan Matusiewicz, Ryan McGarry, Daye Nam, Tiago Cardal Pais, Sheena Panthaplackel, Alaa Saade, Anthony Stewart, Brian Stoler, Mukund Sundararajan, Corentin Tallec, Elaine Thai, Chandu Thekkath, Daniel Toyama, and Roopali Vij.

We also thank everyone who participated in our various user studies and surveys, the entire Learning for Code team at Google DeepMind, and the anonymous reviewers.

\balance
\bibliographystyle{ACM-Reference-Format}
\bibliography{outlines}

\clearpage
\crefalias{section}{appendix}

\appendix

\section{Author Contributions}
\label{app:authors}
This work was led by Kensen Shi and advised by Satish Chandra and Charles Sutton. There were three main workstreams with the following contributors (listed in the same order as in the paper's author list):

\textbf{Core NL outlines:} Kensen Shi, Deniz Altınbüken, Marc Rasi, Brandon Ruffin, Pengcheng Yin, Manzil Zaheer, Satish Chandra, Charles Sutton

\textbf{Android security application:} Kensen Shi, Saswat Anand, Mihai Christodorescu, Roy Tu

\textbf{Code review application:} Kensen Shi, Katja Grünwedel, Alexa Koenings, Sai Naidu, Anurag Pathak, Fredde Ribeiro, Siddhant Sanyam, Maxim Tabachnyk, Sara Toth, Tobias Welp

\section{Taxonomy of Code Explanation}
\label{app:taxonomy}

NL outlines are a novel way of explaining code
with different benefits and drawbacks 
compared to other types of code explanation like docstrings and tutorials.
To clarify the distinctions, we can classify types of code explanations by:
\begin{enumerate}
    \item \textbf{Topic:} It may primarily explain the \emph{interface} of some code, its \emph{implementation}, the \emph{design} decisions it required, details of \emph{APIs and concepts} involved, etc.
    \item \textbf{Audience:} The target readers could be \emph{novice} programmers, \emph{experienced} developers, \emph{maintainers} of the code, \emph{users} of an API, \emph{project managers}, etc.
    \item \textbf{Location:} Explanations can occur \emph{within code} like names and inline comments, in a separate \emph{document} like a design document or bug report, within \emph{developer tools} like IDEs or code search interfaces, in \emph{Web sites} like Javadocs, \emph{ephemerally} in conversation with an LLM chatbot, etc.
    \item \textbf{Length:} Effective explanations can be as short as a \emph{single symbol} like a well chosen method name, or much longer like a \emph{sentence}, \emph{paragraph}, or even a \emph{book}.
\end{enumerate}

NL outlines occupy a particular space within this taxonomy. In terms of topic, NL outlines focus on the implementation strategy: the high-level steps taken to achieve the function's goal. The audience comprises experienced developers who wish to write, maintain, or build upon the code. NL outlines would primarily be shown within code and in developer tools. Finally, an outline is a list of sentence-length statements which in total are much shorter than the code they describe.

This helps distinguish NL outlines from other forms of code explanation. For example, docstrings focus on the contract of a function and its usage instead of the implementation~\cite{PEP257}, and are more often read by API users than code maintainers. Pseudocode is more literal than NL outlines (e.g., explicitly providing every variable update instead of abstracting out the broad concepts) and usually appears in textbooks~\cite{CLRS} and rarely in professional software projects. LLM explanations also vary widely depending on the user's prompt~\cite{GILT}, but they are often ephemeral and long explanations going into great detail to help novices and developers unfamiliar with the concepts involved, unlike the concise format of NL outlines which persist in the code or developer tooling. Design documents are a longer form of code explanation, aiming to describe a software architecture including constraints on the design and alternatives that were considered \cite{ER2023}, ultimately presenting a different set of information for a different audience (e.g., project managers) compared to NL outlines.

At first glance, NL outlines may seem similar to inline code comments, especially because NL outlines may be \emph{displayed} like comments.
Interestingly, \citet{UNDERSTANDABILITY} measured how code understandability correlates to various metrics including the presence of comments, but ultimately obtained a negative result: the number of comments was not correlated with perceived or actual understanding. We take this as evidence that the style of comments typically written by developers is not optimized for code understanding---developers usually write sporadic comments about pitfalls, developer intent, and reminders for future work~\cite{CLASSIFYING-COMMENTS,TOTT-COMMENTS}, without summarizing the functionality more completely. In contrast, our NL outlines focus on the high-level ideas in the function's implementation, are written in a way that still makes sense when presented without code, and are designed to boost code understanding (e.g., overwhelmingly rated as ``Very helpful'' or ``Extremely helpful'' in \cref{fig:outline-quality-survey,fig:security-helpfulness}).

\section{\emph{Finish Changes} with Multiple Changes}
\label{app:finish-changes}

\begin{figure*}
    \centering
    \includegraphics[width=\textwidth]{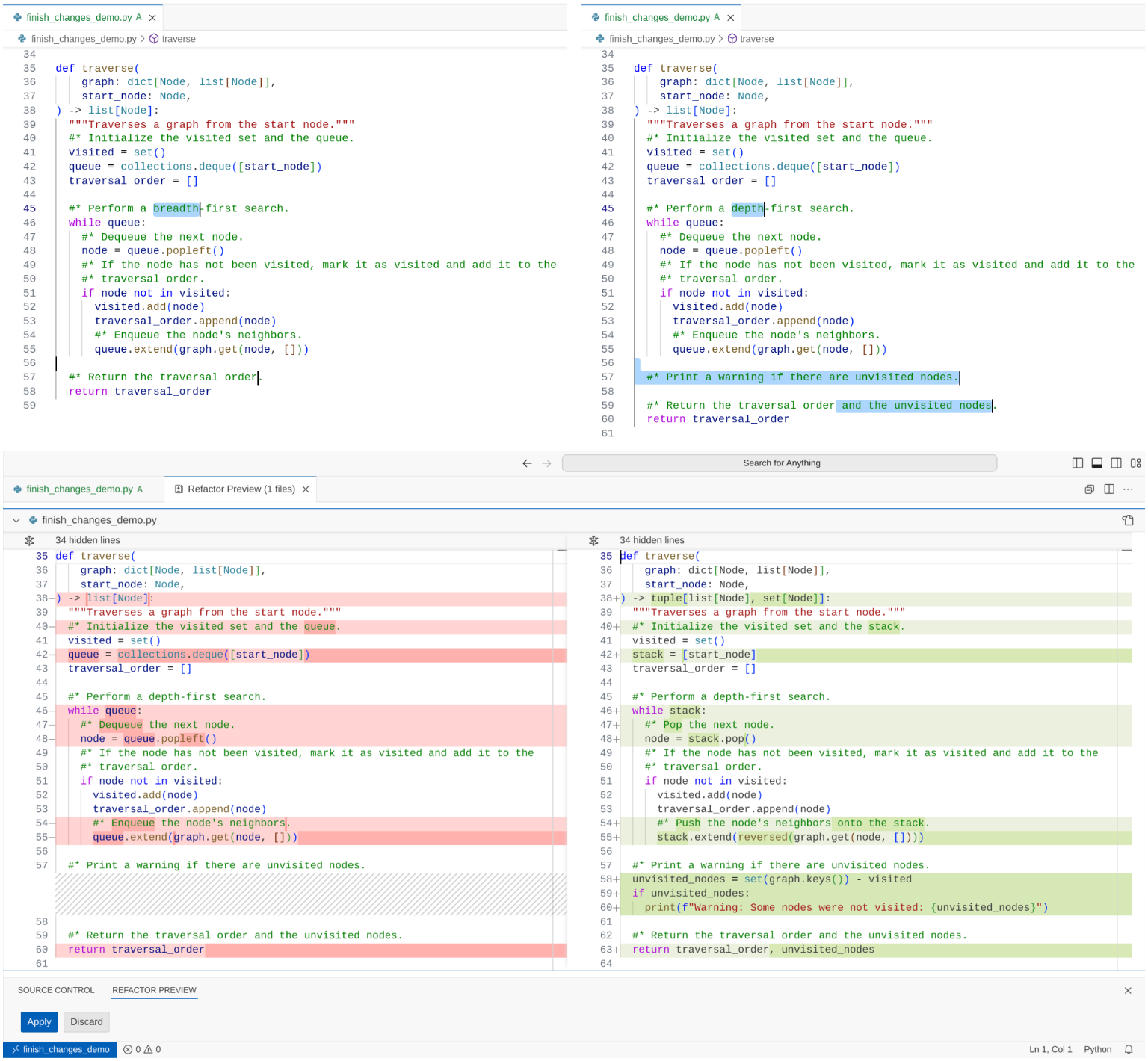}
    \caption{\emph{Top left:} An initial function for which Gemini 1.5 Flash predicts an NL outline that correctly describes a BFS graph traversal. The user will make edits to the outline at the three cursor locations shown. \enskip\emph{Top right:} The user makes three separate changes to the outline by typing the highlighted text, which includes a completely new outline statement. \enskip\emph{Bottom:} Upon invoking the \emph{Finish Changes} feature, the LLM must simultaneously update the traversal algorithm from breadth-first to depth-first search, compute the unvisited nodes and possibly print a warning, and alter the return statement and type annotation. Despite the complex request, Gemini 1.5 Flash correctly suggests numerous updates to the code and outline to finish the user's three changes simultaneously.}
    \label{fig:finish-changes-appendix}
    \vspace{10mm}  
\end{figure*}

\cref{sec:use-cases} describes using NL outlines for code maintenance, in particular suggesting a \emph{Finish Changes} feature which allows the LLM to finish any changes the user has made to the code or outline. \cref{fig:finish-changes} illustrates two example usages of the \emph{Finish Changes} feature, each involving one conceptual edit. But actually, Gemini 1.5 Flash is powerful enough to handle multiple changes at once: \cref{fig:finish-changes-appendix} shows a successful implementation of three simultaneous changes.

In our experience, the \emph{Finish Changes} feature with Gemini 1.5 Flash can suggest surprisingly good edits, but it is not always perfect. The likelihood of slight mistakes tends to increase with more changes requested simultaneously, or if one change requires propagating many different kinds of edits to different locations. We observed that the most common mistake is an incomplete propagation of changes that lacks some necessary downstream edit. But, this also means that most of the suggested changes are on the right track.

Gemini 1.5 Flash occasionally suggests an ``improvement'' that is not strictly needed to finish the user's changes, such as adding more detail to a docstring which was not made stale in the first place. The actual suggestions are usually very reasonable, but this behavior may be distracting to users.

We chose to explore the \emph{Finish Changes} feature with Gemini 1.5 Flash due to its low latency---we receive predicted diffs in about 5 seconds, allowing developers to iterate on \mbox{code quickly}.

\section{Constrained Generation}
\label{app:constrained-generation}

Recall from \cref{sec:generation} that Interleaved Generation has the drawback of potentially changing the code against our wishes, resulting in predicted outlines that no longer apply to the original code. Our Constrained Generation approach resolves this issue by using constrained decoding~\cite{FST,GUIDED-GENERATION}: we construct a regex-like constraint that forces the model to repeat the code exactly, except with optional comment lines between code lines. If the original code parses, we can also require an outline statement before the first line of the function body (after the function signature and docstring) and prevent outline statements from being added above a blank line or in the middle of a multi-line statement or comment block. These extra heuristics help ensure that outline statements are placed appropriately. We found that constraints with these heuristics are broad enough to permit all ``ground-truth'' outlines that we hand-wrote for prompting and experimentation, while being precise enough to avoid almost all formatting mistakes from Interleaved Generation. However, the downside of Constrained Generation is in the implementation and portability: only certain LLMs work with certain constrained decoding APIs, and some constrained decoding engines only support constraints up to a maximum size which limits the length of functions we can outline (but a workaround to these constraint size limitations is to use a sequence of constrained decodings, each continuing where the previous one left off).

\section{Example Prompts for Generating NL Outlines}
\label{app:prompts}

We provide example LLM prompts illustrating the Interleaved Generation and Line Number Infilling techniques from \cref{sec:generation} which were used in the experiments in \cref{sec:experiments}. In the actual experiments, we used few-shot prompting with 8 hand-crafted examples. The prompts below each contain a different subset of those examples for brevity. Long lines are wrapped where indicated with the \mbox{\textcolor{red}{\ensuremath{\hookrightarrow}\space}} symbol. Some LLMs accept system instructions and conversations with multiple turns between the user and LLM. When applicable, we structure the prompt in that way according to the \code{SYSTEM INSTRUCTIONS:}, \code{USER:}, and \code{ASSISTANT:} delimiters. For other LLMs, we send the prompt as one block of text including those delimiters.

Below is an Interleaved Generation prompt with 2 few-shot examples, where \code{\textcolor{red!70!black}{\string{\string{\string{code\_to\_be\_outlined\string}\string}\string}}} is a placeholder for the code for which we want to generate an outline:

\begin{lstlisting}[style=CommonStyle,basicstyle=\fontencoding{T1}\ttfamily\scriptsize,aboveskip=8pt,belowskip=8pt,escapechar=~]
SYSTEM INSTRUCTIONS:
You are an expert programmer.
You are especially good at understanding and explaining the main ideas in a code function.
Your task is to write comments that summarize the main ideas in the code.

Follow these rules:
* Use the comments to organize the code into logical sections.
* Do not change the code aside from adding comments.
* Do not remove or change any existing comments. Only add comments.
* Each comment should be one sentence or phrase.
* When applicable, the comment should explain why the code is written that way, but only if the reasoning is unclear.
* The comment should not be too detailed, so it is quick to read.
* Do not add any comments to the docstring.
* Aim for at most 3 comments for short functions, or at most 5 comments for long functions.
* Do not comment every line.
* Do not explain in words. Only provide the code with comments added, nothing else.

USER:
Please help me understand this code:
```
def download_kaggle_dataset(
    api: kaggle.KaggleApi,
    user: str,
    dataset_name: str,
    target_folder: Path,
    dataset_version: Optional[int] = None,
) -> None:
  """Download a Kaggle dataset and extracts contents to the target folder."""
  target_folder.mkdir(parents=True, exist_ok=True)

  res: response.HTTPResponse = api.datasets_download(
      user, dataset_name,
      dataset_version_number=dataset_version,
      _preload_content=False,
  )

  file_name = dataset_name + '.zip'
  with (target_folder / file_name).open('wb') as f:
    f.write(res.data)

  res.close()

  with zipfile.ZipFile(target_folder / file_name, 'r') as f:
    f.extractall(target_folder)

  os.remove(target_folder / file_name)
```

Identify the logical sections of the code and summarize them by adding comments. Only provide the code with comments, nothing else.

ASSISTANT:
```
def download_kaggle_dataset(
    api: kaggle.KaggleApi,
    user: str,
    dataset_name: str,
    target_folder: Path,
    dataset_version: Optional[int] = None,
) -> None:
  """Download a Kaggle dataset and extracts contents to the target folder."""
  # Create the target folder if needed.
  target_folder.mkdir(parents=True, exist_ok=True)

  # Use the Kaggle API to download the desired dataset as a zip file.
  res: response.HTTPResponse = api.datasets_download(
      user, dataset_name,
      dataset_version_number=dataset_version,
      _preload_content=False,
  )

  # Save the downloaded zip file to the target folder.
  file_name = dataset_name + '.zip'
  with (target_folder / file_name).open('wb') as f:
    f.write(res.data)

  res.close()

  # Extract the dataset contents from the zip file.
  with zipfile.ZipFile(target_folder / file_name, 'r') as f:
    f.extractall(target_folder)

  # Delete the zip file, so that only the extracted contents remain.
  os.remove(target_folder / file_name)
```

USER:
Please help me understand this code:
```
def _get_line_number_prefix(line_number: int) -> str:
  if line_number >= 1000:
    raise ValueError(f'The line number {line_number} is too big.')
  if line_number <= 0:
    # TODO: handle negative numbers?
    raise ValueError(f'The line number must be positive: {line_number}')
  # TODO: turn the width and the separator (currently a tab) into parameters
  return f'{line_number:>3d}\t'
```

Identify the logical sections of the code and summarize them by adding comments. Only provide the code with comments, nothing else.

ASSISTANT:
```
def _get_line_number_prefix(line_number: int) -> str:
  # Check that line_number is positive and not too large.
  if line_number >= 1000:
    raise ValueError(f'The line number {line_number} is too big.')
  if line_number <= 0:
    # TODO: handle negative numbers?
    raise ValueError(f'The line number must be positive: {line_number}')
  # Return a fixed-length string containing the line number.
  # TODO: turn the width and the separator (currently a tab) into parameters
  return f'{line_number:>3d}\t'
```

USER:
Please help me understand this code:
```
~\textcolor{red!70!black}{\string{\string{\string{code\_to\_be\_outlined\string}\string}\string}}~
```

Identify the logical sections of the code and summarize them by adding comments. Only provide the code with comments, nothing else.

ASSISTANT:
\end{lstlisting}

When the LLM responds with text, we assume that it follows the format as demonstrated by the few-shot examples, enabling us to programmatically parse the response to extract the added line comments which form the NL outline.

Below is a similar prompt for the Line Number Infilling technique using 3 different few-shot examples. Similar to before, we use the placeholder \code{\textcolor{red!70!black}{\string{\string{\string{code\_with\_line\_numbers\string}\string}\string}}} for the code we want to generate an outline for, except with line numbers prepended in the same way as in the few-shot examples:

\begin{lstlisting}[style=CommonStyle,basicstyle=\fontencoding{T1}\ttfamily\scriptsize,aboveskip=8pt,belowskip=8pt,escapechar=~]
SYSTEM INSTRUCTIONS:
You are an expert programmer.
You are especially good at understanding and explaining the main ideas in a code function.
Your task is to write comments that summarize the main ideas in the code.

Follow these rules:
* First, write the line number where a logical section of the code starts.
* Then, write a comment to explain and summarize that section of the code.
* Write only one comment for each logical section of the code.
* Each comment should be one sentence or phrase.
* When applicable, the comment should explain why the code is written that way, but only if the reasoning is unclear.
* The comment should not be too detailed, so it is quick to read.
* Do not add any comments to the docstring.
* Aim for at most 3 comments for short functions, or at most 5 comments for long functions.
* Do not comment every line.
* Do not repeat the code in your response.

USER:
Please help me understand this code, with line numbers added for reference:
```
  1|def random_new_variable(existing_variables: List[str],
  2|                        ordered: bool) -> str:
  3|  """Returns a new variable token not in existing_variables."""
  4|  if ordered:
  5|    for i in range(dsl.MAX_NUM_VARIABLES):
  6|      v = dsl.variable_token(i)
  7|      if v not in existing_variables:
  8|        return v
  9|    raise ValueError('Could not find new variable.')
 10|  else:
 11|    choices = list(dsl.ALL_VARIABLES - set(existing_variables))
 12|    if not choices:
 13|      raise ValueError('Could not find new variable.')
 14|    return random.choice(choices)
```

Identify the logical sections of the code and summarize them. Format your response by providing, for each logical section, the line number where that section starts, followed by one sentence that summarizes that section.

Do not repeat the code! Just provide the summary.

ASSISTANT:
4| If ordered, return the first variable that is not in existing_variables.
10| Otherwise, return a variable chosen at random among those not in existing_variables.

USER:
Please help me understand this code, with line numbers added for reference:
```
  1|  @parameterized.named_parameters(
  2|      ('true', True, ['tf.add(in1, in2)', 'tf.add(in2, in1)']),
  3|      ('false', False, ['tf.add(in1, in2)',
  4|                        'tf.add(in2, in1)',
  5|                        'tf.add_n((in1, in2))',
  6|                        'tf.add_n((in2, in1))']))
  7|  def test_run_value_search_only_minimal_solutions(
  8|      self, only_minimal_solutions, expected_solutions):
  9|    benchmark = benchmark_module.Benchmark(
 10|        examples=[benchmark_module.Example(inputs=[[1, 4], [2, 7]],
 11|                                           output=[3, 11])])
 12|    results = value_search.run_value_search(
 13|        benchmark=benchmark,
 14|        settings=settings_module.from_dict({
 15|            'timeout': 20,
 16|            'max_solutions': 4,
 17|            'only_minimal_solutions': only_minimal_solutions,
 18|            'max_extra_solutions_time': 20}))
 19|    self.assertLen(results.solutions, len(expected_solutions))
 20|    self.assertEqual([solution.expression for solution in results.solutions],
 21|                     expected_solutions)
```

Identify the logical sections of the code and summarize them. Format your response by providing, for each logical section, the line number where that section starts, followed by one sentence that summarizes that section.

Do not repeat the code! Just provide the summary.

ASSISTANT:
9| Create a benchmark to run on.
12| Run the value search with different settings for only_minimal_solutions.
19| Check that the resulting solutions are as expected.

USER:
Please help me understand this code, with line numbers added for reference:
```
~\textcolor{red!70!black}{\string{\string{\string{code\_with\_line\_numbers\string}\string}\string}}~
```

Identify the logical sections of the code and summarize them. Format your response by providing, for each logical section, the line number where that section starts, followed by one sentence that summarizes that section.

Do not repeat the code! Just provide the summary.

ASSISTANT:
\end{lstlisting}

We can similarly parse the LLM's response programmatically to extract the location and content of each outline statement, which combine to form the NL outline prediction.

\section{Errors when Parsing Outline Predictions}
\label{app:errors}

\cref{sec:experiments,tab:outline-stats} mention errors detected while parsing an LLM's outline predictions. Here we provide more details about the specific errors and their frequency.

Here is a list of all errors we checked for when parsing an outline prediction using the Interleaved Generation technique, including the severity (major or minor), the frequency in the experiment (number of LLM predictions where the error occurred, out of 150 predictions considering the 5 models and 30 functions), a description of the error's meaning, and what we do when encountering the error:
\begin{itemize}[leftmargin=10pt]
    \item \textbf{Extra blank line (minor, 19 predictions)}: When performing a top-down line-by-line comparison of the predicted code with outline versus the original code (i.e., searching for added comments which form the interleaved outline), we encounter a blank line in the prediction where the corresponding line in the original code is not blank (lines containing only whitespace are also considered blank). We simply skip the blank line in the prediction, as if it did not exist.
    \item \textbf{Missing blank line (minor, 7)}: When performing the same line-by-line comparison as described above, we encounter a blank line in the original code where the corresponding line in the prediction is not blank. We similarly skip this line in the original code, as if the prediction did contain a matching blank line.
    \item \textbf{Consecutive comment (major, 0)}: We encounter multiple consecutive comment lines present in the prediction but not in the original code. We concatenate the text of the comments into one outline statement, just in case the model decided to split a long comment onto multiple lines, but the long comment length likely indicates poor quality.
    \item \textbf{Missing comment (minor, 20)}: We encounter a comment line in the original code where the prediction does not have a corresponding match. We simply skip this line in the original code, as if the prediction did contain a matching comment.
    \item \textbf{Changed trailing comment (minor, 1)}: We encounter corresponding lines in the prediction and original code that do not match, and neither is a blank line or comment line, and they \emph{do} match if we ignore trailing comments. We simply move on, as if the prediction's trailing comment did match the original code's trailing comment (if any).
    \item \textbf{Changed code (major, 10)}: We encounter corresponding lines in the prediction and original code that do not match, and none of the prior cases apply. We stop processing the prediction and return any outline statements already parsed, because any further outline statements are for predicted code that has deviated significantly from the original code.
    \item \textbf{Extra prediction lines (minor, 0)}: We matched all lines in the original code to corresponding lines in the prediction, but the prediction still has unmatched lines remaining. We ignore those lines, as if the prediction stopped in accordance with the original code.
    \item \textbf{Missing prediction lines (major, 0)}: We matched all lines in the prediction to corresponding lines in the original code, but the original code still has unmatched lines remaining. The predicted outline is likely incomplete since the prediction stopped early, but there is not much we can do about it (without another LLM query).
    \item \textbf{Empty outline (major, 2)}: After the parsing process, the outline contains zero outline statements. This is clearly a poor prediction, but there is not much we can do about it.
\end{itemize}

Although the Constrained Generation technique is not used in our experiments (\cref{sec:experiments}), its predictions are formatted like those for Interleaved Generation. Hence, we use the same procedure to parse the LLM predictions and detect errors (although some are no longer possible due to the constraints).

Here is an analogous list of errors we checked for when using the Line Number Infilling technique:
\begin{itemize}[leftmargin=10pt]
    \item \textbf{Malformed line (major, 1 prediction)}: The prediction line does not follow the expected format of line number, separator, and outline statement. We skip this line.
    \item \textbf{Line number out of bounds (major, 0)}: The line number is smaller than 1 or greater than the number of lines in the original code. We skip this line.
    \item \textbf{Not sorted (minor, 0)}: The prediction lines were not sorted by line number in ascending order. We sort the lines.
    \item \textbf{Duplicate line number (major, 0)}: The prediction contained multiple lines with the same line number. We keep the first instance and ignore the latter duplicates.
    \item \textbf{Commented empty line (minor, 6)}: A line number actually points to a blank line in the original code. Since it is unnatural to add a comment above a blank line (the blank line should go above the comment, which is then directly above its section of code), we increase the line number such that the outline statement is placed above the next non-empty line.
    \item \textbf{Empty outline (major, 0)}: After the parsing process, the outline contains zero outline statements. This is clearly a poor prediction, but there is not much we can do about it.
\end{itemize}

\section{Quality Survey Questions}
\label{app:survey-questions}

\cref{sec:experiments} and \cref{fig:outline-quality-survey} refer to a survey we used to gather human opinions about NL outline quality. In total, we obtained $300$ outline predictions: $30$ functions in the dataset $\times$ $5$ LLMs $\times$ $2$ generation techniques. For each function, we asked the contributor of that function to inspect the function's $10$ predicted outlines, which were presented in shuffled order without any labeling of which LLM or technique produced that outline. For each outline, the function contributor answered five survey questions with multiple-choice answers, listed below. Survey participants were also given a document containing clarifications to the questions, also listed below.

\begin{enumerate}[leftmargin=14pt]
    \item How good is the outline \textbf{overall}, relative to an optimal outline for this function?\\
    $\rightarrow$ Excellent / Good / Acceptable / Bad / Terrible
    \begin{itemize}
        \item This question measures the room for improvement in the LLM predictions.
        \item Don't judge the broad concept of NL Outlines, or that outlines may be fundamentally less useful for certain kinds of functions. Instead, judge the outline sample relative to the best outline you can think of for that function. How close is this outline to being the \emph{best it can be} for that function?
        \item The factors in the other questions (helpfulness, correctness, detail, style/fluency) may factor into your judgment for this question. How much to consider each factor is up to you, so that you can judge the outline holistically.
        \item As a point of reference, a completely empty outline should be considered ``Bad''. Note that a ``Terrible'' outline can be worse than this, if it ends up being a big waste of time (e.g., a lot of words that are obviously nonsense, or a convincing but incorrect outline that leads to future bugs).
    \end{itemize}

    \item How \textbf{helpful} would this outline be in practice?\\
    $\rightarrow$ Very helpful / Somewhat helpful / Not helpful
    \begin{itemize}
        \item This question measures the practical utility of the outline, relative to the baseline of not having outlines at all.
        \item How helpful would this outline be to developers in general, including the code author and other developers who may have varying amounts of familiarity with the code?
        \item Consider the purpose of NL Outlines and multiple potential use-cases, as described above.
        \item You may consider that some functions (e.g., very simple functions) might naturally benefit less from outlines.
        \item Assume that readers of the outline understand that it is AI-generated and may be incorrect, and they act reasonably with that understanding. The amount of trust they have in the outline may depend on how convincing the outline seems, within reason.
    \end{itemize}

    \item Is the outline factually \textbf{correct}?\\
    $\rightarrow$ Completely correct / Mostly correct / Incorrect
    \begin{itemize}
        \item Does the outline say factually correct things?
        \item Use your expertise of the function and its context within the broader project.
        \item It is not necessary to be 100\% perfect here. Think about how you would summarize the function to a colleague to help them gain an initial high-level understanding. You don't want to confuse your colleague with misinformation, but if some detail would take a paragraph to fully hash out, you'd likely do your best to give a ``good enough'' explanation instead, striking a balance between exhaustive correctness and efficient explanation.
        \item Omitting important information is primarily a ``Needs more detail'' issue (for the question below), but it may also affect this correctness question if the missing information leaves the reader with a factual misunderstanding (like ``lying by omission'').
    \end{itemize}

    \item Does the outline have the right \textbf{amount of detail}?\\
    $\rightarrow$ Right amount of detail / Needs more detail / Too much detail
    \begin{itemize}
        \item Does the outline help you (the author \emph{or other developers}) understand the code \emph{more quickly}? Omitting important details, or including too many details, can both slow down understanding.
        \item Use your expertise of the function and how it plays a role in the larger project, to determine which details are important enough to say. What should another developer know if they wanted to find a bug or implement a new feature, in this function or a related one?
        \item It is possible for an outline to simultaneously include unnecessary details, while also missing important details. In this case, choose ``Needs more detail'' or ``Too much detail'' depending on which is the bigger problem for this outline.
    \end{itemize}

    \item How is the outline's \textbf{style and fluency}? Consider formatting, grammar, tone, clarity, etc.\\
    $\rightarrow$ Excellent / Good / Bad
    \begin{itemize}
        \item Is the language clear or ambiguous? Is the outline easy to misunderstand?
        \item Are you distracted by awkward phrasing or grammatical errors?
        \item Are the outline statements placed correctly? If a statement is slightly misplaced, that doesn't necessarily impact the factual correctness or amount of detail, but should be considered here.
        \item Some predictions may seem unnaturally cut off. Indeed this may happen if the model's prediction becomes malformed somehow. That kind of issue should be reflected here, and in the other questions as appropriate.
    \end{itemize}

\end{enumerate}

\section{Example NL Outlines}
\label{app:example-outlines}

\cref{fig:outline-example-main}, \cref{fig:outline-example-test}, \cref{fig:outline-example-restrict-memory}, \cref{fig:outline-example-get-file-explanation-good}, and \cref{fig:outline-example-parse-prediction-good} showcase a variety of high-quality NL outlines for different functions, using various LLMs and generation techniques. We also show examples with commentary about broader patterns:
Gemini 1.0 Pro often gives too much detail (\cref{fig:outline-example-get-file-explanation-too-much-detail}),
Gemini 1.5 Pro sometimes gives too little detail (\cref{fig:outline-example-parse-prediction-needs-more-detail}), and Line Number Infilling is the more difficult technique leading to certain common issues with weaker LLMs (\cref{fig:outline-example-parse-prediction-wrong}). In all examples, we redacted names and edited whitespace to avoid awkward line breaks.

\newpage\clearpage
\nobalance

\begin{figure}[H]
\small
LLM: \textbf{Gemini 1.5 Pro}, Technique: \textbf{Interleaved Generation} \smallskip

Survey responses: $\bigstar$\,\emph{Excellent} (overall), $\bigstar$\,\emph{Very helpful}, $\bigstar$\,\emph{Completely correct}, $\bigstar$\,\emph{Right amount of detail}, $\bigstar$\,\emph{Excellent} (style) \smallskip

    \begin{lstlisting}[style=MyPython,escapechar=~]
def main(argv):
  ~\highlight{\#*\normalfont{ Check for correct usage and set random seed.}}~
  if len(argv) > 1:
    raise app.UsageError('Too many command-line arguments.')

  np.random.seed(FLAGS.seed)
  ~\highlight{\#*\normalfont{ Generate random data for the TSP instances.}}~
  data = np.random.uniform(
      size=(FLAGS.dataset_size, FLAGS.graph_size, 2))
  per_instance_seeds = np.random.randint(
      1000000, size=(FLAGS.dataset_size))

  ~\highlight{\#*\normalfont{ Select the TSP solver based on the given flag.}}~
  if FLAGS.solver == NEAREST_NEIGHBOR:
    solver = nearest_neighbor
  elif FLAGS.solver == FARTHEST_INSERTION:
    solver = farthest_insertion
  elif FLAGS.solver == FARTHEST_INSERTION_SAMPLING:
    solver = functools.partial(farthest_insertion_sampling,
                               num_samples=FLAGS.num_samples,
                               unique_samples=FLAGS.unique_samples,
                               temperature=FLAGS.temperature,
                               caching=FLAGS.caching)
  elif FLAGS.solver == FARTHEST_INSERTION_BS:
    solver = functools.partial(farthest_insertion_bs,
                               num_samples=FLAGS.num_samples,
                               temperature=FLAGS.temperature)
  else:
    raise app.UsageError(
        'Unknown solver: {}'.format(FLAGS.solver))

  ~\highlight{\#*\normalfont{ Run the solver on each instance and measure the total time.}}~
  start_time = timeit.default_timer()
  solutions = []
  for instance, seed in zip(data, per_instance_seeds):
    np.random.seed(seed)
    solutions.append(solver(instance))
  elapsed_time = timeit.default_timer() - start_time

  ~\highlight{\#*\normalfont{ Compute the total cost and check the validity of solutions.}}~
  dataset_cost = 0
  for cost, tour in solutions:
    if sorted(tour) != list(range(FLAGS.graph_size)):
      raise ValueError('Tour is malformed.')
    dataset_cost += cost

  ~\highlight{\#*\normalfont{ Print the results and settings used.}}~
  print('Dataset size: {}'.format(FLAGS.dataset_size))
  print('Graph size: {}'.format(FLAGS.graph_size))
  print('Seed: {}'.format(FLAGS.seed))
  print('Solver: {}'.format(FLAGS.solver))
  if FLAGS.solver == FARTHEST_INSERTION_SAMPLING:
    print()
    print('Sampling-related options:')
    print('  Num samples: {}'.format(FLAGS.num_samples))
    print('  Unique samples: {}'.format(FLAGS.unique_samples))
    print('  Temperature: {}'.format(FLAGS.temperature))
    print('  Caching: {}'.format(FLAGS.caching))
  if FLAGS.solver == FARTHEST_INSERTION_BS:
    print()
    print('Beam search options:')
    print('  Num samples: {}'.format(FLAGS.num_samples))
    print('  Temperature: {}'.format(FLAGS.temperature))
  print()
  print('Time: {:.2f} sec'.format(elapsed_time))
  print('Average cost: {:.5f}'.format(dataset_cost / len(data)))
\end{lstlisting}
    \caption{A high-quality NL outline predicted by Gemini 1.5 Pro using the Interleaved Generation technique, which was rated the best option ($\bigstar$) for each dimension surveyed. The outline statements are very helpful in accurately summarizing what happens in each section of code. Interestingly, the LLM correctly describes ``TSP instances'' and ``the TSP solver'' even though the code does not explicitly say ``TSP'' or ``Traveling Salesman Problem'' (and no prompt example does either). The LLM uses its broad knowledge to infer what is going on, likely using clues from the code such as the different solver names and the ``graph'' and ``tour'' terminology.}
    \label{fig:outline-example-main}
\end{figure}

\begin{figure}[H]
\small
LLM: \textbf{Gemini 1.5 Pro}, Technique: \textbf{Line Number Infilling} \smallskip

Survey responses: $\bigstar$\,\emph{Excellent} (overall), $\bigstar$\,\emph{Very helpful}, $\bigstar$\,\emph{Completely correct}, $\bigstar$\,\emph{Right amount of detail}, $\bigstar$\,\emph{Excellent} (style) \smallskip

    \begin{lstlisting}[style=MyPython,escapechar=~]
@parameterized.named_parameters(
    ('with_description', 'do foo to tensor', True),
    ('empty_description', '', False),
    ('none_description', None, False))
@mock.patch('sys.stdout', new_callable=six.StringIO)
def test_add_constants_and_inputs_and_print_handles_description(
    self, description, expect_print, mock_stdout):
  ~\highlight{\#*\normalfont{ Create a benchmark with one example and a description.}}~
  examples = [benchmark_module.Example(inputs=[[1, 4], [2, 7]],
                                       output=[3, 11])]
  benchmark = benchmark_module.Benchmark(
      examples, description=description)
  ~\highlight{\#*\normalfont{ Call the function being tested.}}~
  values_by_weight = [collections.OrderedDict()
                      for _ in range(5)]
  output_value = value_module.OutputValue(
      benchmark.examples[0].output)
  value_search._add_constants_and_inputs_and_print(
      values_by_weight, benchmark, output_value,
      self._constant_operation, self.settings)
  ~\highlight{\#*\normalfont{ Check that the description was printed to stdout if and only if it was non-empty.}}~
  if expect_print:
    self.assertIn('Description: {}\n'.format(description),
                  mock_stdout.getvalue())
  else:
    self.assertNotIn('Description:', mock_stdout.getvalue())
\end{lstlisting}
    \caption{A high-quality NL outline predicted by Gemini 1.5 Pro using Line Number Infilling. This outline helps the reader understand the structure and intent of the test. The first outline statement describes \code{benchmark} which is the actual object of interest, constructed with a varying \code{description} in different test cases. The second outline statement correctly implies that \code{values\_by\_weight} and \code{output\_value} are created only to ``call the function being tested.'' The last outline statement correctly describes the intent, i.e., to check that the description is printed ``if and only if it was non-empty'' which is consistent with the parameterized test cases. The LLM impressively identifies this intent without a docstring or any comments in the original code.}
    \label{fig:outline-example-test}
\end{figure}

\begin{figure}[H]
\small
LLM: \textbf{Gemini 1.5 Flash}, Technique: \textbf{Interleaved Generation} \smallskip

Survey responses: $\bigstar$\,\emph{Excellent} (overall), $\bigstar$\,\emph{Very helpful}, $\bigstar$\,\emph{Completely correct}, $\bigstar$\,\emph{Right amount of detail}, $\bigstar$\,\emph{Excellent} (style) \smallskip

    \begin{lstlisting}[style=MyPython,escapechar=~]
@contextlib.contextmanager
def restrict_memory(bytes_limit: int):
  """Context manager to restrict memory usage."""
  ~\highlight{\#*\normalfont{ Get the current memory limits.}}~
  soft, hard = resource.getrlimit(resource.RLIMIT_AS)
  ~\highlight{\#*\normalfont{ Set the new soft limit to the minimum of the current soft limit and the given limit.}}~
  new_soft = min(soft, bytes_limit) if soft > 0 else bytes_limit
  resource.setrlimit(resource.RLIMIT_AS, (new_soft, hard))
  ~\highlight{\#*\normalfont{ Execute the code within the context.}}~
  try:
    yield
  ~\highlight{\#*\normalfont{ Restore the original memory limits.}}~
  finally:
    resource.setrlimit(resource.RLIMIT_AS, (soft, hard))
\end{lstlisting}
    \caption{A high-quality NL outline predicted by Gemini 1.5 Flash using Interleaved Generation. This outline helps the reader interpret the code, which is especially helpful because this function uses uncommon mechanisms (implementing a context manager) and APIs (e.g., \code{resource.getrlimit}, \code{resource.setrlimit}, and \code{resource.RLIMIT\_AS}).}
    \label{fig:outline-example-restrict-memory}
\end{figure}

\begin{figure}[H]
\small
LLM: \textbf{Gemini 1.5 Pro}, Technique: \textbf{Line Number Infilling} \smallskip

Survey responses: $\bigstar$\,\emph{Excellent} (overall), $\bigstar$\,\emph{Very helpful}, $\bigstar$\,\emph{Completely correct}, $\bigstar$\,\emph{Right amount of detail}, $\bigstar$\,\emph{Excellent} (style) \smallskip

    \begin{lstlisting}[style=MyPython,escapechar=~]
def get_file_explanation(
    request: redacted.FileExplanationRequest,
    model: llm_lib.ModelSpec,
) -> redacted.FileExplanationResponse:
  """Gets an explanation for each function in a file."""
  ~\highlight{\#*\normalfont{ Record the start time and attempt to extract functions from the given code.}}~
  start_time = timeit.default_timer()
  try:
    functions = structured_function_utils.get_functions(
        request.code)
  except Exception as e: # pylint: disable=broad-exception-caught
    return redacted.FileExplanationResponse(
        success=False,
        error_message=str(e),
    )

  ~\highlight{\#*\normalfont{ Define a helper function to get an explanation for a single function.}}~
  def _explain_function(line_number):
    function_request = redacted.FunctionExplanationRequest(
        code=request.code,
        file_path=request.file_path,
        line=line_number,
        column=1,
    )
    return get_function_explanation(function_request, model)

  ~\highlight{\#*\normalfont{ Get explanations for all functions in the code using multiple threads.}}~
  function_line_numbers = [
      line_number for _, line_number in functions
  ]
  num_threads = min(10, len(functions))
  with multiprocessing.pool.ThreadPool(num_threads) as pool:
    explanations = list(
        pool.map(_explain_function, function_line_numbers))

  ~\highlight{\#*\normalfont{ Construct the response object, populating it with the function explanations.}}~
  response = redacted.FileExplanationResponse(success=True)
  for function_explanation in explanations:
    response.function_explanations.append(function_explanation)

  ~\highlight{\#*\normalfont{ Log the elapsed time and return the response.}}~
  elapsed_time = timeit.default_timer() - start_time
  logging.info(
      'Got file explanation (%d functions) in %.1f seconds.',
      len(explanations),
      elapsed_time,
  )

  return response
\end{lstlisting}
    \caption{A high-quality NL outline predicted by Gemini 1.5 Pro using Line Number Infilling. This outline has particularly effective wording that efficiently communicates important details. For example, ``attempt to extract functions'' alludes to the \code{try}/\code{except} structure without necessitating a separate outline statement for the \code{except} block, and ``using multiple threads'' is an important detail that readers should know about.}
    \label{fig:outline-example-get-file-explanation-good}
\end{figure}

\begin{figure}[H]
\small
LLM: \textbf{Gemini 1.0 Pro}, Technique: \textbf{Interleaved Generation} \smallskip

Survey responses: \xmark\,\emph{Bad} (overall), $\pmb{\sim}$\,\emph{Somewhat helpful}, $\bigstar$\,\emph{Completely correct}, \xmark\,\emph{Too much detail}, $\pmb{\sim}$\,\emph{Good} (style) \smallskip

    \begin{lstlisting}[style=MyPython,escapechar=~]
def get_file_explanation(
    request: redacted.FileExplanationRequest,
    model: llm_lib.ModelSpec,
) -> redacted.FileExplanationResponse:
  """Gets an explanation for each function in a file."""
  ~\highlight{\#*\normalfont{ Start a timer to measure how long it takes to get the file explanation.}}~
  start_time = timeit.default_timer()
  ~\highlight{\#*\normalfont{ Try to extract functions from the code.}}~
  try:
    functions = structured_function_utils.get_functions(
        request.code)
  ~\highlight{\#*\normalfont{ If there is an error, return an error response.}}~
  except Exception as e: # pylint: disable=broad-exception-caught
    return redacted.FileExplanationResponse(
        success=False,
        error_message=str(e),
    )

  ~\highlight{\#*\normalfont{ Define a function to explain a single function.}}~
  def _explain_function(line_number):
    ~\highlight{\#*\normalfont{ Create a function explanation request.}}~
    function_request = redacted.FunctionExplanationRequest(
        code=request.code,
        file_path=request.file_path,
        line=line_number,
        column=1,
    )
    ~\highlight{\#*\normalfont{ Get the explanation for the function.}}~
    return get_function_explanation(function_request, model)

  ~\highlight{\#*\normalfont{ Get the line numbers of all the functions in the file.}}~
  function_line_numbers = [
      line_number for _, line_number in functions
  ]
  ~\highlight{\#*\normalfont{ Use a thread pool to explain the functions in parallel.}}~
  num_threads = min(10, len(functions))
  with multiprocessing.pool.ThreadPool(num_threads) as pool:
    explanations = list(
        pool.map(_explain_function, function_line_numbers))

  ~\highlight{\#*\normalfont{ Create a file explanation response.}}~
  response = redacted.FileExplanationResponse(success=True)
  ~\highlight{\#*\normalfont{ Add the explanations to the response.}}~
  for function_explanation in explanations:
    response.function_explanations.append(function_explanation)

  ~\highlight{\#*\normalfont{ Log the time it took to get the file explanation.}}~
  elapsed_time = timeit.default_timer() - start_time
  logging.info(
      'Got file explanation (%d functions) in %.1f seconds.',
      len(explanations),
      elapsed_time,
  )

  ~\highlight{\#*\normalfont{ Return the response.}}~
  return response
\end{lstlisting}
    \caption{A low-quality NL outline predicted by Gemini 1.0 Pro for the same function as in \cref{fig:outline-example-get-file-explanation-good}. Although this outline is entirely correct, it has too much detail and is \emph{inefficient} at conveying the important ideas, lowering the outline's helpfulness and overall quality. For comparison, the outline in \cref{fig:outline-example-get-file-explanation-good} is much better. This example reflects a more general pattern where Gemini 1.0 Pro often provides too much detail.}
    \label{fig:outline-example-get-file-explanation-too-much-detail}
\end{figure}

\begin{figure}[H]
\small
LLM: \textbf{Gemini 1.0 Pro}, Technique: \textbf{Interleaved Generation} \smallskip

Survey responses: $\bigstar$\,\emph{Excellent} (overall), $\bigstar$\,\emph{Very helpful}, $\bigstar$\,\emph{Completely correct}, $\bigstar$\,\emph{Right amount of detail}, $\bigstar$\,\emph{Excellent} (style) \smallskip

    \begin{lstlisting}[style=MyPython,escapechar=~]
def parse_prediction(
    prediction: str,
    prompt_strategy: str = 'summary_verdict_outline',
    summary_line_width: int | None = None,
) -> ParsedPrediction:
  """Extracts (summary, suspicion score, outline, errors) from
  the response."""
  ~\highlight{\#*\normalfont{ Handle the case where the prediction contains extra text.}}~
  if '\n\n\n' in prediction:
    prediction = prediction.split('\n\n\n')[0]
  errors = []

  ~\highlight{\#*\normalfont{ Parse the prediction based on the prompt strategy.}}~
  if prompt_strategy == 'summary_verdict_outline':
    ~\highlight{\#*\normalfont{ Check for missing sections and return an error if any are found.}}~
    if '\n\nSuspicion score:\n' not in prediction:
      errors.append('[no score section]')
    if '\n\nNotes:\n' not in prediction:
      errors.append('[no outline section]')
    if errors:
      return ParsedPrediction(summary='', score=-1,
                              outline=[], errors=errors)
    ~\highlight{\#*\normalfont{ Extract the summary, score, and outline text from the prediction.}}~
    summary, _, remaining = prediction.partition(
        '\n\nSuspicion score:\n')
    score, _, outline_text = remaining.partition('\n\nNotes:\n')

  elif prompt_strategy == 'outline_summary_verdict':
    raise ValueError('outline_summary_verdict is deprecated.')
  else:
    raise ValueError(
        f'Unhandled prompt strategy: {prompt_strategy}')

  ~\highlight{\#*\normalfont{ Clean up the summary, score, and outline text.}}~
  summary = summary.strip()
  score = score.strip()
  if score in ['0', '1', '2', '3']:
    score = int(score)
  else:
    errors.append(f'[unexpected score: {score}]')
    score = -1
  outline_text = outline_text.strip()

  ~\highlight{\#*\normalfont{ Parse the outline text into a list of tuples containing line numbers and comments.}}~
  outline = []
  if outline_text != '<None>':
    for line in outline_text.splitlines():
      match_result = re.fullmatch(
          r'^Lines? (\d+)(?: ?- ?\d+)?: (.*)$', line)
      if not match_result:
        errors.append('[malformed outline line]')
        continue
      line_number = int(match_result.group(1))
      comment = match_result.group(2)
      outline.append((line_number, comment))

  ~\highlight{\#*\normalfont{ Sort the outline by line number and check for duplicate line numbers.}}~
  outline = sorted(outline, key=lambda pair: pair[0])
  i = 1
  while i < len(outline):
    this_line = outline[i][0]
    last_line = outline[i-1][0]
    if this_line == last_line:
      errors.append(f'[duplicate line number: {this_line}]')
      outline.pop(i)
    else:
      i += 1

  ~\highlight{\#*\normalfont{ Wrap the summary lines to the specified width, if provided.}}~
  summary_lines = summary.splitlines()
  if summary_line_width:
    summary = '\n'.join(
        ['\n'.join(textwrap.wrap(line, width=summary_line_width))
         for line in summary_lines])

  ~\highlight{\#*\normalfont{ Return a ParsedPrediction object containing the summary, score, outline, and errors.}}~
  return ParsedPrediction(summary=summary, score=score,
                          outline=outline, errors=errors)
\end{lstlisting}
    \caption{Gemini 1.0 Pro predicts an excellent outline for the same function as in \cref{fig:outline-example-parse-prediction-needs-more-detail} and \cref{fig:outline-example-parse-prediction-wrong}. Although Gemini 1.0 Pro often includes \emph{too much} detail, for this function the extra detail is very helpful (without being excessive as in \cref{fig:outline-example-get-file-explanation-too-much-detail}).}
    \label{fig:outline-example-parse-prediction-good}
\end{figure}

\begin{figure}[H]
\small
LLM: \textbf{Gemini 1.5 Pro}, Technique: \textbf{Interleaved Generation} \smallskip

Survey responses: $\bigstar$\,\emph{Excellent} (overall), $\bigstar$\,\emph{Very helpful}, $\bigstar$\,\emph{Completely correct}, \xmark\,\emph{Needs more detail}, $\bigstar$\,\emph{Excellent} (style) \smallskip

    \begin{lstlisting}[style=MyPython,escapechar=~]
def parse_prediction(
    prediction: str,
    prompt_strategy: str = 'summary_verdict_outline',
    summary_line_width: int | None = None,
) -> ParsedPrediction:
  """Extracts (summary, suspicion score, outline, errors) from
  the response."""
  ~\highlight{\#*\normalfont{ Remove any text after two consecutive empty lines.}}~
  if '\n\n\n' in prediction:
    prediction = prediction.split('\n\n\n')[0]
  errors = []

  ~\highlight{\#*\normalfont{ Extract the summary, score, and outline based on the prompt strategy.}}~
  if prompt_strategy == 'summary_verdict_outline':
    if '\n\nSuspicion score:\n' not in prediction:
      errors.append('[no score section]')
    if '\n\nNotes:\n' not in prediction:
      errors.append('[no outline section]')
    if errors:
      return ParsedPrediction(summary='', score=-1,
                              outline=[], errors=errors)
    summary, _, remaining = prediction.partition(
        '\n\nSuspicion score:\n')
    score, _, outline_text = remaining.partition('\n\nNotes:\n')

  elif prompt_strategy == 'outline_summary_verdict':
    raise ValueError('outline_summary_verdict is deprecated.')
  else:
    raise ValueError(
        f'Unhandled prompt strategy: {prompt_strategy}')

  ~\highlight{\#*\normalfont{ Parse the score and outline.}}~
  summary = summary.strip()
  score = score.strip()
  if score in ['0', '1', '2', '3']:
    score = int(score)
  else:
    errors.append(f'[unexpected score: {score}]')
    score = -1
  outline_text = outline_text.strip()

  outline = []
  if outline_text != '<None>':
    for line in outline_text.splitlines():
      match_result = re.fullmatch(
          r'^Lines? (\d+)(?: ?- ?\d+)?: (.*)$', line)
      if not match_result:
        errors.append('[malformed outline line]')
        continue
      line_number = int(match_result.group(1))
      comment = match_result.group(2)
      outline.append((line_number, comment))

  ~\highlight{\#*\normalfont{ Check for duplicate line numbers in the outline.}}~
  outline = sorted(outline, key=lambda pair: pair[0])
  i = 1
  while i < len(outline):
    this_line = outline[i][0]
    last_line = outline[i-1][0]
    if this_line == last_line:
      errors.append(f'[duplicate line number: {this_line}]')
      outline.pop(i)
    else:
      i += 1

  ~\highlight{\#*\normalfont{ Wrap the summary lines if needed.}}~
  summary_lines = summary.splitlines()
  if summary_line_width:
    summary = '\n'.join(
        ['\n'.join(textwrap.wrap(line, width=summary_line_width))
         for line in summary_lines])

  return ParsedPrediction(summary=summary, score=score,
                          outline=outline, errors=errors)
\end{lstlisting}
    \caption{An example outline where Gemini 1.5 Pro is correct but does not include enough detail. Specifically, the outline here fails to describe error handling, the content of \code{outline}, that \code{outline} is sorted, and the returned object. In contrast, \cref{fig:outline-example-parse-prediction-good} shows a better outline for the same function with more detail, predicted by Gemini 1.0 Pro. In our experience, lack of detail is the most common issue in outlines predicted by Gemini 1.5 Pro (albeit a relatively minor issue, since such outlines have the benefit of being very quick to read). Performing LLM-specific prompt engineering would allow us to encourage more detail from Gemini 1.5 Pro.}
    \label{fig:outline-example-parse-prediction-needs-more-detail}
\end{figure}

\begin{figure}[H]
\small
LLM: \textbf{DeepSeek-Coder-Instruct}, Technique: \textbf{Line Number Infilling}\smallskip

Survey responses: \xmark\,\emph{Terrible} (overall), \xmark\,\emph{Not helpful}, \xmark\,\emph{Incorrect}, \xmark\,\emph{Too much detail}, $\pmb{\sim}$\,\emph{Good} (style) \smallskip

    \begin{lstlisting}[style=MyPython,escapechar=~]
def parse_prediction(
    prediction: str,
    prompt_strategy: str = 'summary_verdict_outline',
    summary_line_width: int | None = None,
) -> ParsedPrediction:
  ~\highlight{\#*\normalfont{ Extract summary, suspicion score, outline, and errors from the prediction.}}~
  """Extracts (summary, suspicion score, outline, errors) from
  the response."""
  if '\n\n\n' in prediction:
    prediction = prediction.split('\n\n\n')[0]
  errors = []

  ~\highlight{\#*\normalfont{ Handle the \textquotesingle summary\_verdict\_outline\textquotesingle{} prompt strategy.}}~
  if prompt_strategy == 'summary_verdict_outline':
    if '\n\nSuspicion score:\n' not in prediction:
      errors.append('[no score section]')
    if '\n\nNotes:\n' not in prediction:
      errors.append('[no outline section]')
    if errors:
      return ParsedPrediction(summary='', score=-1,
                              outline=[], errors=errors)
    ~\highlight{\#*\normalfont{ Extract summary, suspicion score, and outline from the prediction.}}~
    summary, _, remaining = prediction.partition(
        '\n\nSuspicion score:\n')
    score, _, outline_text = remaining.partition('\n\nNotes:\n')

  ~\highlight{\#*\normalfont{ Handle the \textquotesingle outline\_summary\_verdict\textquotesingle{} prompt strategy.}}~
  elif prompt_strategy == 'outline_summary_verdict':
    raise ValueError('outline_summary_verdict is deprecated.')
  else:
    raise ValueError(
        f'Unhandled prompt strategy: {prompt_strategy}')

  summary = summary.strip()
  score = score.strip()
  if score in ['0', '1', '2', '3']:
    score = int(score)
  else:
    errors.append(f'[unexpected score: {score}]')
    score = -1
  outline_text = outline_text.strip()

  ~\highlight{\#*\normalfont{ Validate the score and format the summary.}}~
  outline = []
  if outline_text != '<None>':
    for line in outline_text.splitlines():
      match_result = re.fullmatch(
          r'^Lines? (\d+)(?: ?- ?\d+)?: (.*)$', line)
      if not match_result:
        errors.append('[malformed outline line]')
        continue
      line_number = int(match_result.group(1))
      comment = match_result.group(2)
      outline.append((line_number, comment))

  ~\highlight{\#*\normalfont{ Extract outline from the prediction.}}~
  outline = sorted(outline, key=lambda pair: pair[0])
  i = 1
  while i < len(outline):
    this_line = outline[i][0]
    last_line = outline[i-1][0]
    if this_line == last_line:
      errors.append(f'[duplicate line number: {this_line}]')
      outline.pop(i)
    else:
      i += 1

  ~\highlight{\#*\normalfont{ Format the summary.}}~
  summary_lines = summary.splitlines()
  if summary_line_width:
    summary = '\n'.join(
        ['\n'.join(textwrap.wrap(line, width=summary_line_width))
         for line in summary_lines])

  ~\highlight{\#*\normalfont{ Return the parsed prediction.}}~
  return ParsedPrediction(summary=summary, score=score,
                          outline=outline, errors=errors)
\end{lstlisting}
    \caption{A low-quality outline predicted by DeepSeek-Coder-Instruct using Line Number Infilling for the same function as in \cref{fig:outline-example-parse-prediction-good} and \cref{fig:outline-example-parse-prediction-needs-more-detail}. Line Number Infilling is the more difficult technique, especially for weaker LLMs. Common issues include outline statements that are misplaced (e.g., above the docstring), missing (e.g., above \code{summary = summary.strip()}), or wrong (e.g., ``Validate the score and format the summary.'').}
    \label{fig:outline-example-parse-prediction-wrong}
\end{figure}

\section{Virtual Split for Code Review}
\label{app:virtual-split}

\cref{subsec:code-review} discusses the Virtual Split feature which helps code reviewers tackle complex change lists (CLs). Here we provide illustrative screenshots, lightly edited for clarity and space.

\begin{figure*}
    \centering
    \includegraphics[width=0.75\textwidth,trim=0mm 48mm 0mm 68mm,clip]{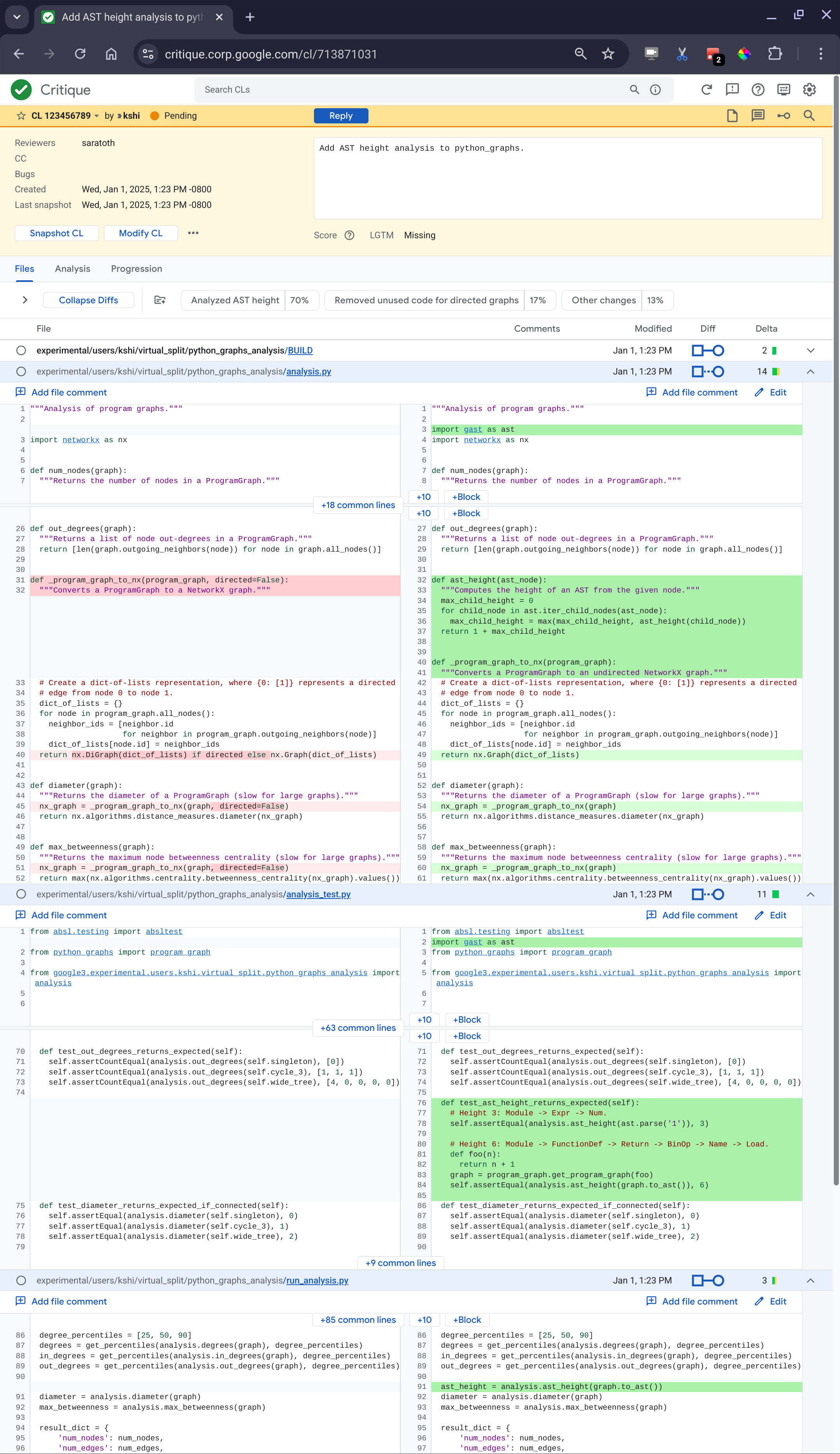}
    \vspace{-1mm}
    \caption{An example CL shown in Critique, Google's internal code review platform. Critique shows CL metadata, a list of topics provided by the Virtual Split feature, and the list of changed files with their diffs and metadata.}
    \label{fig:virtual-split-no-topic}
\end{figure*}

\cref{fig:virtual-split-no-topic} shows an example CL in Critique, Google's internal code review platform. The top area contains metadata about the CL, including its author, any assigned reviewers, and the CL description provided by the CL author (in this case, ``Add AST height analysis to \texttt{python\_graphs}'').
Then, a diff navigation toolbar is provided, containing a ``Expand Diffs'' or ``Collapse Diffs'' button followed by series of buttons representing topics in the Virtual Split. Each topic button contains a short description of the topic (e.g., ``Analyzed AST height'') and the percentage of the CL's changed lines which are assigned to that topic. The initial state is shown where none of the topic buttons have been clicked yet.
Finally, a table is shown where each row contains metadata about one changed file in the CL. Each row is clickable and displays or collapses that file's diff. In this case, the diff for \code{BUILD} was left collapsed, while diffs for \code{analysis.py}, \code{analysis\_test.py}, and \code{run\_analysis.py} have been expanded.

\begin{figure*}
    \centering
    \includegraphics[width=0.75\textwidth,trim=0mm 53mm 0mm 68mm,clip]{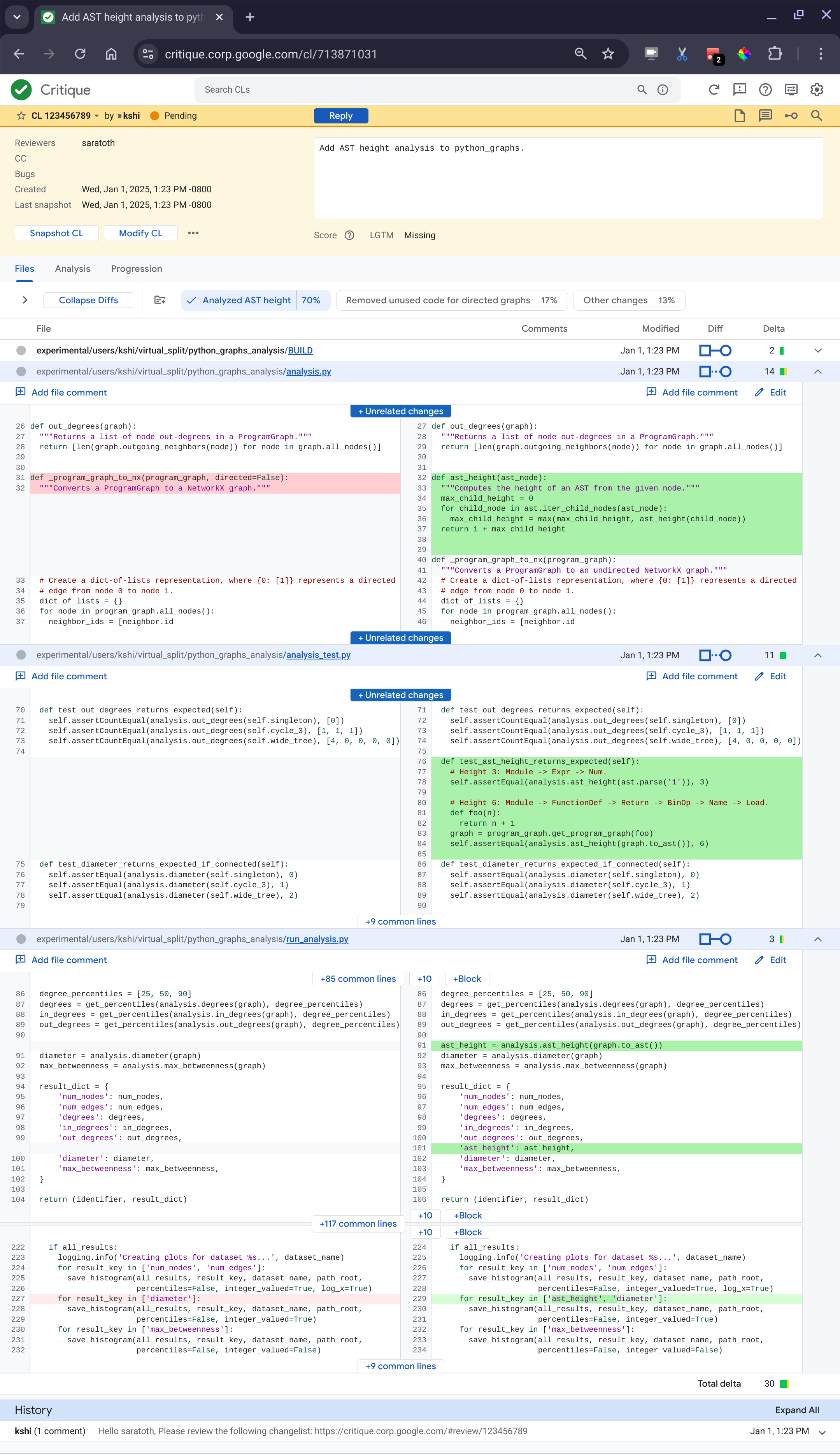}
    \vspace{-2mm}
    \caption{After clicking the first topic button for ``Analyzed AST height'', only the relevant parts of diffs are expanded, with unrelated changes collapsed. This helps reviewers focus on one conceptual change and reduces scrolling between tightly-coupled changes in different files (in this case, the \code{ast\_height} function's implementation, tests, and usage).}
    \label{fig:virtual-split-topic-1}
\end{figure*}

Without Virtual Split, a CL reviewer would not see any topic buttons, and they would typically read each file diff from top to bottom, either in the file order listed on screen or using the reviewer's intuitions for which files to read first. However, users report that it is difficult to quickly identify the most important changes to review, and to simultaneously keep in mind all of the changes happening within a complex CL (often much more complex than this example).
\cref{fig:virtual-split-topic-1} shows the state after the user clicks the first topic button, i.e., ``Analyzed AST height''. This automatically expands only the portions of diffs related to that topic, collapsing unrelated changes in the file (e.g., the \code{\_program\_graph\_to\_nx} changes in \code{analysis.py}) or entire unrelated files (e.g., \code{BUILD}). This way, reviewers can focus on changes within this topic without being distracted by unrelated changes. For example, the reviewer can see the new \code{ast\_height} function in \code{analysis.py}, its unit tests in \code{analysis\_test.py}, and its usage in \code{run\_analysis.py}, all adjacent to each other to minimize scrolling and context switching. Users still have the option to expand any collapsed diffs to see more context as needed.

\begin{figure*}
    \centering
    \includegraphics[width=0.75\textwidth,trim=0mm 551mm 0mm 68mm,clip]{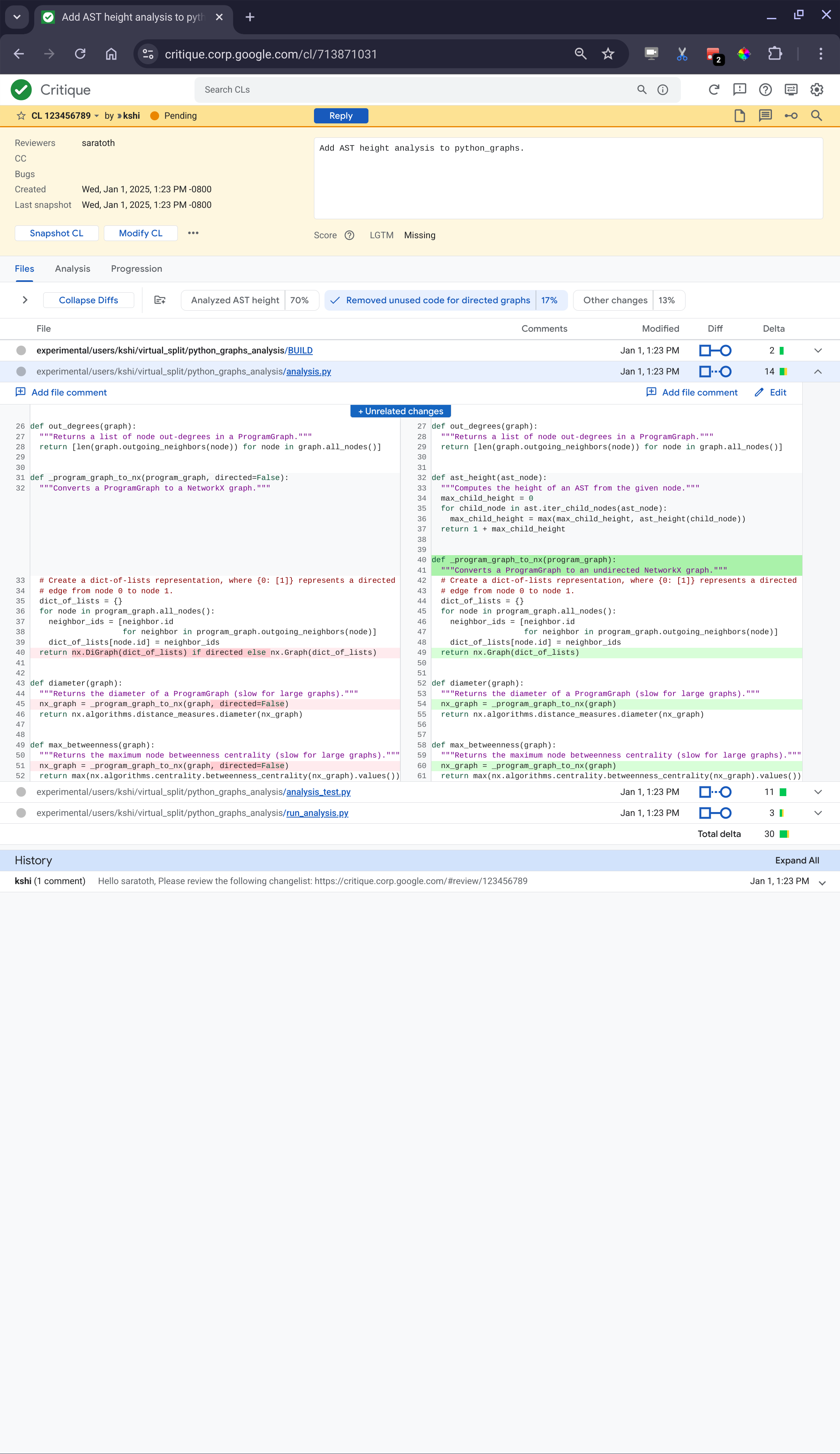}
    \caption{The second topic, ``Removed unused code for directed graphs'', shows changes to the \code{\_program\_graph\_to\_nx} function and its calls. The LLM also correctly predicts that these changes are less important than the AST height analysis (the first topic). Thus, the ordering of topics can help reviewers better understand the CL intention and prioritize their review effort. The last topic, ``Other changes'', is designed to encapsulate changes that are low-priority for review, including imports, typo fixes, formatting changes, and so on.}
    \label{fig:virtual-split-topic-2}
\end{figure*}

\cref{fig:virtual-split-topic-2} shows the state after the user clicks the second topic button, ``Removed unused code for directed graphs''. This expands changes to the \code{\_program\_graph\_to\_nx} function signature, implementation, and calls in \code{analysis.py}. Even though some unrelated changes in that file are visible within the displayed diff context, those unrelated changes are highlighted in gray instead of the usual red/green highlighting. Furthermore, since \code{analysis\_test.py} is not expanded, the reviewer can then infer that the \code{\_program\_graph\_to\_nx} changes do not come with associated changes to unit tests (assuming that the Virtual Split is accurate).

Our system guarantees that every changed part of every file is assigned to some topic, so that once a reviewer has clicked through all of the topic buttons, they have seen all of the changes in the CL. Furthermore, the LLM tries to order the topics from most to least important, thus helping reviewers quickly identify and navigate to the most salient parts of the CL. Overall, the topic descriptions help reviewers understand a complex CL at a glance, and the topic buttons guide reviewers through an organized, focused, and sequentially prioritized viewing of the changes.


\section{Drawbacks to Star Comments}
\label{app:star-comments-downsides}

Star comments, proposed in \cref{sec:discussion}, have important downsides in churn, flexibility, and distraction.

\emph{Churn:} Source code is expensive to store and edit considering the costs of version-controlled storage forever, the human cost of code review, and extra strain on continuous testing infrastructure.\footnote{Adding comments can unfortunately cause tests to fail. Detecting and fixing these issues across a large codebase is a real cost.} Putting all iterations of NL outlines (which change alongside the code or through LLM upgrades or experimentation) into source code is thus surprisingly expensive. Alternatively, storing NL outlines as non-code metadata in a central database or in extra files adjacent to the source code (but without version control, mandatory code review, or continuous testing) would significantly reduce churn.

\emph{Flexibility:} Star comments reduce the flexibility to update outlines without causing more churn. Even if the code is not changed, it may still be desirable to update or deprecate outlines due to prompting improvements, LLM upgrades, or human-provided feedback, and to perform A/B experiments providing users with different outlines. These would be straightforward with outlines stored as non-code metadata, but with star comments, updates are expensive due to churn and A/B experiments are impossible.

\emph{Distraction:} Star comments in the code are by default visible, like any other code. This may be distracting to users who suddenly see outlines in developer tooling that does not yet support enough outline-related features to make them worthwhile. Developers need the ability to hide outlines at times, so this functionality must be built into various tooling at the outset. Alternatively, outlines as metadata would be hidden by default and can be gradually shown to users depending on their enthusiasm (e.g., beta testers) and the tooling readiness.

In our view, star comments may be used to explore using NL outlines in practice with a limited scope of users and files (e.g., a small team or small company), but outlines should be stored separately from code in the long term at large companies.

\section{Generalizing Outlines to Files and Projects}
\label{app:generalization}

This paper focuses on NL outlines for individual functions because code flow within a function is naturally structured: code flows from top down, excluding control flow jumps. This implies that adjacent code lines are likely to be semantically related, forming clusters of related lines. Hence, an NL outline is designed to \emph{partition} the lines into those logical clusters.

This is no longer true at the level of a class or file, where code flow is less predictable without enforcement of an organized ordering of functions or classes within a file. \emph{If} functions were well-organized, then it would be straightforward to partition them into semantic groups with outline statements, but it is less clear what should be done if the functions are not organized. Should the LLM suggest a reordering? That could help keep files more organized but could also confuse developers unaware of the reordering. Alternatively, we may create a file outline that is not aligned with the code (and thus cannot be shown interleaved with the code), e.g., an unordered mapping of functions to semantic groups. The list of symbols in a file (provided in an IDE sidebar) typically allows the user to sort the symbols by code position or alphabetically, and this idea may suggest a new semantically-grouped ordering where each group is summarized with an outline statement.

Outlining a project may mean generation of a README file that summarizes the main ideas, lists important directories and files and their purpose, and provides examples of usage or application flow. It may also include generated diagrams illustrating the user journeys, data flows, class hierarchies, and so on. The outline may incorporate human-written sections and could be automatically updated as the code evolves.

We imagine that outlining a file or project likely begins with generating NL outlines for functions first, and then using those for further LLM prompting, reasoning, abstraction, and summarization in a hierarchical manner, and combined with static or dynamic analyses where applicable. In this sense, NL outlines can be building blocks in complex LLM pipelines.

\end{document}